\documentclass[particles,article,accept,moreauthors,pdflatex]{Definitions/mdpi} 
\usepackage{amssymb,color,graphicx,bm,mathrsfs,color,amsmath,slashed,comment}
\usepackage{amsmath}
\usepackage{hyperref}
\newcommand{\be}{\begin{equation}}
\newcommand{\ee}{\end{equation}}
\newcommand{\bea}{\begin{eqnarray}}
\newcommand{\eea}{\end{eqnarray}}
\newcommand{\half}{\frac1 2}

\newcommand{\vectau}{{\bm \tau}}
\newcommand{\vecrho}{{\bm \rho}}
\newcommand{\ie}{{\it i.e.}}
\newcommand{\eg}{{\it e.g.}}
\newcommand{\MeV}{{\rm MeV}}
\input{acronym.input}
\firstpage{500} 
\pubvolume{3}
\issuenum{2}
\articlenumber{34}
\pubyear{2020}
\copyrightyear{2020}
\history{Received: 22 May 2020; Accepted: 16 June 2020; Published: 19 June 2020}

\pdfoutput=1

\Title{Bulk Viscous Damping of Density Oscillations in Neutron Star Mergers}

\Author{Mark Alford $^{1}$\orcidA{}, 
Arus Harutyunyan $^{2,3}$*\orcidB{} and 
Armen Sedrakian $^{4,5}$\orcidC{}}

\AuthorNames{Mark Alford, Arus Harutyunyan and Armen Sedrakian}

\address{%
$^{1}$ \quad Department of Physics, Washington University, St.~Louis,
  Missouri 63130, USA; alford@physics.wustl.edu
\\
$^{2}$ \quad Byurakan Astrophysical Observatory, National Academy of
Sciences,   Byurakan 0213, Armenia
\corres{Correspondence: arus@bao.sci.am }
\\
$^{3}$ \quad Yerevan State University, Alek Manukyan str. 1,
  Yerevan 0025, Armenia\\
$^{4}$ \quad Frankfurt Institute for Advanced Studies, D-60438
  Frankfurt am Main, Germany; sedrakian@fias.uni-frankfurt.de
\\  
$^{5}$ \quad Institute of Theoretical Physics, University of Wroc\l{}aw,
50-204 Wroc\l{}aw, Poland   }
\corres{Correspondence: arus@bao.sci.am }

\abstract{ In this paper, we discuss the damping of density oscillations in dense nuclear matter in the temperature range relevant to neutron star mergers. This damping is due to bulk viscosity arising from the weak interaction ``Urca'' processes of neutron decay and electron capture. The nuclear matter is modelled in the relativistic density functional approach.  The bulk viscosity reaches a resonant maximum close to the neutrino trapping temperature, then drops rapidly as temperature rises into the range where neutrinos are trapped in neutron stars. We investigate the bulk viscous dissipation timescales in a post-merger object and identify regimes where these timescales are as short as the characteristic timescale $\sim$10 ms, and, therefore, might affect the evolution of the post-merger object. Our analysis indicates that bulk viscous damping would be important at not too high temperatures of the order of a few MeV and densities up to a few times saturation density.}


\begin{document}
\section{Introduction}
The recent detections of gravitational waves by the LIGO-Virgo collaboration, in particular, the multimessenger binary-neutron star (BNS) merger event GW170817~\cite{Abbott2017}, motivate studies of the transport properties of dense nuclear matter at temperatures and densities relevant to BNS
mergers~\cite{Alford2018a,Harutyunyan:2016a,Harutyunyan:2016b,Harutyunyan2018b,Alford2019a,Alford2019b}. The mass of the post-merger object typically would exceed the maximum mass of a neutron star and, as a consequence, it would collapse to a black hole on the timescales ranging from tens of milliseconds up to seconds depending on the mass of the post-merger object~\cite{Faber2012:lrr,Baiotti2017,Baiotti2019}. While gravitational waves in the post-merger phase have not been observed in the GW170817 event due to lack of detector sensitivity at high frequencies,
numerical relativity studies of BNS mergers in their highly non-linear regime predict intense emission of gravitational waves in the kHz frequency range during the initial phase of post-merger phase lasting typically 10 ms~(for recent simulations see, for example, \cite{Endrizzi2018,Most2019,Ciolfi2019,Tsokaros2019}). Improvements at least by a factor of 2 compared to advanced LIGO  design sensitivity are necessary to measure the dominant frequency component of the signal of GW170817-like event and by factors of 4-5 to observe sub-dominant features of post-merger signal~\cite{Torres-Rivas2019}.
It is expected that dissipation or damping of matter flows in the merged stars could influence the gravitational waves emitted during the post-merger phase. Recent estimates of the role of the thermal conduction and shear and bulk viscosities indicate that, of these, damping of density oscillations via bulk viscosity has the strongest
influence~\cite{Alford2018a,Alford2019a,Alford2019b}. After a brief introduction to the problem of computation of the bulk viscosity, we present an extension of our recent work~\cite{Alford2019b} on the bulk viscosity of nucleonic matter which includes the estimates of timescales of the damping of the oscillations by the bulk viscosity. The relevance of the bulk viscosity will be assessed by comparing the damping timescales of density oscillations to the characteristic timescales of the initial phase of post-merger
 $\sim 10$ ms (over which the post-merger object is expected to emit intense gravitational waves) as well as the longer-term phase  $\sim 1$ s.
 
The bulk viscosity of nuclear matter at temperatures up to about 1\,MeV has been studied extensively~\cite{Sawyer1979ApJ,Sawyer1980ApJ,Sawyer1989,Haensel1992PhRvD,Haensel:2000vz,Haensel:2001mw,Haensel:2001em,Dong2007,Alford2010JPhG,Alford:2010jf,Kolomeitsev2015} in the context of oscillations of neutron stars and, in particular, as a source of damping of (unstable) $r$-mode oscillations; for reviews see~\cite{Schmitt2017,Kokkotas2016EPJA}. More recently, interest in neutron star mergers has motivated studies of bulk viscosity of nucleonic matter at temperatures up to several tens of MeV, covering both the regime where neutrinos escape from neutron stars and the regime where they are trapped~\cite{Alford2019a,Alford2019b}. Such high temperatures significantly affect the phase space occupation of the fermions and, therefore, the rates of the weak-interaction processes \cite{Alford2018b}. Once the temperature becomes high enough for neutrinos to be trapped, they affect the composition of matter and ensure that direct Urca processes are always kinematically possible, and modified Urca is a subleading correction.

In this contribution, we extend recent work~\cite{Alford2019b} to compute the timescales associated with bulk-viscous damping of oscillations. We use two models of the equation of state (EoS) and associated composition of dense matter which are based on the relativistic density functional theory of nuclear matter. The overall picture is that, at densities from around $n_0$ (nuclear saturation density) to around $3 n_0$, the bulk viscosity reaches a maximum at temperature $T\simeq 2\div 6$\,MeV, which is in the regime where neutrinos are not (or not completely) trapped. At these temperatures, therefore, the  damping timescale is at a minimum, with values ranging from about 10 ms at low densities $n_B\leq n_0$ down to milliseconds (or even tenths of milliseconds depending on the EoS)
 at $n_B\simeq 3n_0$. This means that bulk viscous damping can have noticeable effects during the $\sim 10$\,ms of initial  (gravitational-wave-emission) phase of the post-merger. At higher temperatures where the neutrino-trapping occurs the bulk viscosity falls by orders of magnitude, which implies much longer damping timescales which are larger than the characteristic timescales involved.
  
This paper is organized as follows. In~Section~\ref{sec:bulk} we discuss the
formalism for computing the bulk viscosity and the approximations involved in such a computation. Our focus is on the beta equilibration processes of neutron decay and electron capture and the microscopic relaxation rates associated with these processes. Section~\ref{sec:results} starts with a brief discussion of the thermodynamic properties of nuclear matter derived from density functional theory in Section~\ref{sec:DFT}. Bulk viscosity and the oscillation damping timescale are discussed in Sections~\ref{sec:bulk_visc} and \ref{sec:damping}. Our main results are summarized in Section~\ref{sec:summary}. We use natural units ($\hbar = c = k_B = 1$) and the metric signature $g_{\mu\nu} = \textrm{diag}(1, -1, -1, -1)$.

\section{Urca processes and bulk viscosity}
\label{sec:bulk}

We start with a brief reminder of the bulk viscosity of nuclear matter
composed of neutrons, protons, electrons, muons; for more details see~\cite{Alford2019a,Alford2019b}. For~simplicity, we will neglect the muonic contribution to equilibration rates, although~we include their contribution to static thermodynamic quantities such as~susceptibilities.

Above the trapping temperature $T_{\rm tr}\simeq 5$\,MeV, the neutrino mean-free-path is smaller than the size of a neutron star, so neutrinos are trapped in the merger region.  Under these conditions beta equilibrium is established via neutron decay and electron capture, and their inverse processes
\bea\label{eq:Urca_1}
n\rightleftarrows p + e^-+\bar{\nu}_e \ ,\\
\label{eq:Urca_2}
p + e^- \rightleftarrows n+\nu_e \ .
\eea

In $\beta$-equilibrium the chemical potentials of particles obey the relation
\bea\label{eq:eq_condition}
\mu_p+\mu_e = \mu_n+\mu_{\nu} \ .
\eea
The particle fractions of baryonic matter for any given temperature $T$, baryon number density 
$n_B=n_n+n_p$  and lepton densities $n_{L_l}=n_l+n_{\nu_l}=Y_{L_l}n_B$ 
(since we ignore muon reactions, the lepton fractions $Y_{L_l}$ need to be fixed for each flavor separately) are found by imposing the beta-equilibrium 
condition~\eqref{eq:eq_condition}, a similar condition for muons $\mu_p+\mu_\mu = \mu_n+\mu_{\nu_\mu}$ and the charge neutrality condition $n_p=n_e+n_\mu$. 

At lower temperatures $T\lesssim T_{\rm tr}$ the neutrino mean free path is larger than the 
size of a neutron star so the system is neutrino-transparent. Neutrinos cannot occur in initial 
states, therefore  the reactions~\eqref{eq:Urca_1} and \eqref{eq:Urca_2} proceed only in one direction (from left to right).
To determine the composition of matter in this case we use
the ordinary zero-temperature $\beta$-equilibrium conditions
$\mu_n=\mu_p+\mu_e$ and $\mu_\mu=\mu_e$.  Reference~\cite{Alford2018b}
found that there are significant corrections to these conditions
at $T\gtrsim 1$\,MeV; nevertheless the bulk viscosity in the neutrino transparent regime is not affected significantly~\cite{Alford2019a}.

If the matter is driven out of $\beta$-equilibrium, for~example by
compression and rarefaction, the~left and right-hand sides of
Equation~\eqref{eq:eq_condition} do not balance anymore. The~
deviation from $\beta$-equilibrium is then measured by a quantity
\bea\label{eq:mu_Delta}
\mu_\Delta = \mu_n+\mu_{\nu}-\mu_p-\mu_e.
\eea
As a result, the Urca processes~\eqref{eq:Urca_1} and
\eqref{eq:Urca_2} will go faster in one direction than in the other
until the beta equilibrium of matter is restored.

Consider now small-amplitude density oscillations 
in nuclear matter with a frequency $\omega$.
 The baryon and lepton conservation implies for periodic
 perturbations $\delta n_B(t),\delta n_L(t) \sim e^{i\omega t}$ 
\bea\label{eq:cont_i}
\delta n_i(t) =-\frac{\theta}{i\omega}\, n_{i0},
\quad i=\{B,L\},
\eea 
where $n_{B0}=n_{n0}+n_{p0}$ and $n_{L0}=n_{e0}+n_{\nu 0}$ 
are the unperturbed background densities of baryons and leptons,
and $\theta$ is the divergence of fluid velocity.
The compression and rarefaction of matter implies  
perturbations in particle densities which can be 
separated into instantaneous equilibrium and 
non-equilibrium parts
\bea\label{eq:dens_j}
n_j(t)=n_{j0}+\delta n_j(t), \quad \delta n_j(t)=
\delta n^{\rm eq}_j(t)+\delta n'_j(t),\quad j=\{n,p,e,\nu\},
\eea 
where $n_{j0}$ are the static values of particle densities. The
variations $\delta n^{\rm eq}_j(t)$ stand for the shifts of the
equilibrium state for the instantaneous values of $n_{B}(t)$ and 
$n_{L}(t)$, whereas $\delta n'_j(t)$ are the deviations of the 
particle densities from those equilibrium values. There exist 
two choices of the instantaneous equilibrium state. Below, we 
follow our recent work~\cite{Alford2019b}, for the alternative 
see Ref.~\cite{Huang2010}. We compare below these two approaches 
and explain why they give the same result for the bulk viscosity.

The non-equilibrium perturbations $\delta n'_j(t)$ drag
matter out of chemical equilibrium by leading to a small
chemical potential shift~\eqref{eq:mu_Delta} which can 
be written in terms of particle densities as
\bea\label{eq:delta_mu}
\mu_\Delta(t)=A_{n}\delta n_n(t)+A_{\nu} \delta n_\nu(t) -
A_{p} \delta n_p(t) -A_{e}\delta n_e(t),
\eea 
where $A_n=A_{nn}-A_{pn}$, $A_p=A_{pp}-A_{np}$, 
and $A_e=A_{ee}$, $A_\nu=A_{\nu\nu}$ with
\bea\label{eq:A_j}
A_{ij} = \frac{\partial \mu_i}{\partial n_j}.
\eea 
The off-diagonal elements $A_{np}$ and $A_{pn}$ are non-zero 
because of the cross-species strong interaction between neutrons 
and protons. For small amplitude density oscillations we only need to evaluate the derivatives in Equation~\eqref{eq:A_j} at $\mu_\Delta=0$.

If there were no flavor-changing weak processes, 
then the particle densities would just oscillate 
around their static equilibrium values according to
\bea\label{eq:cont_j}
\frac{\partial}{\partial t} \delta {n}_j^0(t)= 
-\theta n_{j0}\quad\Rightarrow\quad
\delta {n}_j^0(t) = -\frac{\theta}{i\omega}\, n_{j0}.
\eea 
The weak interactions lead to an imbalance between
the rates of direct and inverse Urca processes which 
in the ``subthermal regime'' ($\mu_\Delta\ll T$) can be written 
as~\cite{Sawyer1989,Haensel1992PhRvD,Huang2010}
\bea\label{eq:lambda_def}
\Gamma_\Delta \equiv \Gamma_p-\Gamma_n =\lambda\mu_\Delta,\quad\lambda > 0,
\eea
with $\Gamma_p$ and $\Gamma_n$ being the production rates of protons
and neutrons, respectively. The~production rate~\eqref{eq:lambda_def}
should be added to the right hand sides of Equation~\eqref{eq:cont_j} with
a plus sign for $p,e$ and a minus sign for $n,\nu$, for~example, 
\bea\label{eq:cont_n_weak}
\frac{\partial}{\partial t}\delta n_n(t) = 
-\theta  n_{n0} -\lambda\mu_\Delta(t),\quad etc.
\eea 

Substituting here Equation~\eqref{eq:delta_mu}, exploiting the 
conditions $\delta n_B =\delta n_n +\delta n_p$,
$\delta n_p=\delta n_e+\delta n_\mu$ and
$\delta n_L =\delta n_e+\delta n_\nu$, and using Equation~\eqref{eq:cont_i}
for $n_B$, $n_L$ and an analogous equation for $n_\mu$ 
(as muons are assumed not to participate in any reactions, their 
fraction is conserved) one finds
\bea\label{eq:delta_n}
\delta n_n &=& -\frac{i\omega n_{n0} +\lambda
  (A_{p}+A_e+A_\nu)n_{B0}
-\lambda A_\nu n_{L0}-\lambda(A_e+A_\nu)n_{\mu 0}}
{i\omega+\lambda A}\frac{\theta}{i\omega},\\
\label{eq:delta_p}
\delta n_p &=& -\frac{i\omega n_{p0} 
+\lambda A_{n}n_{B0}+\lambda A_\nu n_{L0}+\lambda(A_e+A_\nu)n_{\mu 0}}
{i\omega+\lambda A}\frac{\theta}{i\omega},\\
\label{eq:delta_e}
\delta n_e &=& -\frac{i\omega n_{e0} 
+\lambda A_{n}n_{B0}+\lambda A_\nu n_{L0}-\lambda(A_n+A_p)n_{\mu 0}}
{i\omega+\lambda A}\frac{\theta}{i\omega},\\
\label{eq:delta_nu}
\delta n_\nu &=& -\frac{i\omega n_{\nu 0}+\lambda
  (A_{n}+A_{p}+A_e)n_{L0} -\lambda A_{n} n_{B0}
  +\lambda(A_n+A_p)n_{\mu 0}}{i\omega+\lambda A}\frac{\theta}{i\omega},
\eea 
with a ``beta-disequilibrium--proton-fraction'' susceptibility given by
\bea\label{eq:A_def}
A=\sum_{i} A_{i} 
=\frac{\partial\mu_n}{\partial n_n}
+\frac{\partial\mu_p}{\partial n_p}
-\frac{\partial\mu_n}{\partial n_p}
-\frac{\partial\mu_p}{\partial n_n}
+\frac{\partial\mu_e}{\partial n_e}
+\frac{\partial\mu_\nu}{\partial n_\nu} 
= \biggl(\dfrac{\partial \mu_\Delta}
{\partial n_n} \biggr)_{\!n_B}.
\eea 
Equations~\eqref{eq:delta_n}--\eqref{eq:delta_nu} 
are the extensions of Eqs.~(37)--(39) of Ref.~\cite{Alford2019b}
as we included non-zero muon density $n_\mu$ here, which
was previously neglected. The final formula for the bulk
viscosity, however, remains the same after this addition.

In the next step we find $\delta n_j^{\rm eq}$ using the 
definition of the instantaneous $\beta$-equilibrium state:
$A_{n}\delta n^{\rm eq}_n+A_\nu\delta n^{\rm eq}_\nu-
A_{p}\delta n^{\rm eq}_p-A_e\delta n^{\rm eq}_e=0$, which gives
\bea\label{eq:delta_n_eq}
\delta n^{\rm eq}_n &=& \frac{-(A_{p}+A_e+A_\nu)n_{B0}
+A_\nu n_{L0} +(A_e+A_\nu)n_{\mu 0}}{A} \frac{\theta}{i\omega},\\
\label{eq:delta_p_eq}
\delta n^{\rm eq}_p &=& -\frac{A_{n} n_{B0}
+ A_\nu n_{L0} +(A_e+A_\nu)n_{\mu 0}}{A}\frac{\theta}{i\omega},\\
\label{eq:delta_e_eq}
\delta n^{\rm eq}_e &=& -\frac{A_{n} n_{B0}
+ A_\nu n_{L0} -(A_n+A_p)n_{\mu 0}}{A}\frac{\theta}{i\omega},\\
\label{eq:delta_nu_eq}
\delta n^{\rm eq}_\nu &=& \frac{-(A_{n}+A_{p}+A_e)n_{L0}
+A_{n} n_{B0}-(A_n+A_p)n_{\mu 0}}{A} \frac{\theta}{i\omega}.
\eea 
Note that these expressions are the solutions of the balance
equations~\eqref{eq:cont_n_weak} in the limit of infinite
relaxation rate $\lambda\to\infty$, which implies necessarily
$\mu_\Delta\to 0$. One can check this also by pushing 
$\lambda$ to infinity directly in general solutions 
\eqref{eq:delta_n}--\eqref{eq:delta_nu} which will reduce
then to Equations~\eqref{eq:delta_n_eq}--\eqref{eq:delta_nu_eq}.

Now the non-equilibrium density perturbations can be 
found according to Equation~\eqref{eq:dens_j} 
\bea\label{eq:delta_j'}
\delta n'_p = \delta n'_e =-\delta n'_n = -\delta n'_\nu
  =\frac{ C}{A(i\omega+\gamma)}\theta,
\eea 
where $\gamma=\lambda A$ has a dimension of frequency and
measures the relaxation rate of particle densities to their
equilibrium values, and
\bea\label{eq:C_def}
C = n_{n0} A_{n}+n_{\nu 0} A_\nu - n_{p0}A_{p} - n_{e0}A_e 
= n_B\biggl(\dfrac{\partial \mu_\Delta}{\partial n_B}
\biggr)_{\!Y_n}
\eea 
is the ``beta-disequilibrium--baryon-density'' susceptibility,
with $Y_n=n_n/n_B$ being the neutron fraction.
The non-equilibrium part of the pressure -- the so-called 
bulk viscous pressure, can be now computed as
\bea\label{eq:Pi}
\Pi =\sum_j
\frac{\partial p}{\partial n_j}\delta n'_j
=\sum_{lj} n_{l0}A_{lj}\delta n'_j,
\eea 
where we used the Gibbs-Duhem relation $dp=sdT+\sum_l n_l d \mu_l$.
Substituting the solutions~\eqref{eq:delta_j'} in Equation~\eqref{eq:Pi} 
one finds the bulk viscosity from the definition $\Pi=-\zeta\theta$ 
\bea\label{eq:zeta}
\zeta = \frac{C^2}{A}\frac{\gamma}{\omega^2+\gamma^2}.
\eea
The susceptibility prefactor $C^2/A$ is a pure 
thermodynamic quantity and depends only on the EoS, 
whereas the relaxation rate $\gamma=\lambda A$ 
[Equation~\eqref{eq:lambda_def}] depends on the 
microscopic scattering amplitudes of weak interactions. It is seen from Equations~\eqref{eq:lambda_def} and 
\eqref{eq:A_def} that $\gamma$ is actually the derivative
\bea\label{eq:gamma}
\gamma =\biggl(\dfrac{\partial \Gamma_\Delta}
{\partial \mu_\Delta} \biggr)
\biggl(\dfrac{\partial \mu_\Delta}
{\partial n_n} \biggr)_{\!n_B} 
= \biggl(\dfrac{\partial \Gamma_\Delta}
{\partial n_n} \biggr)_{\!n_B} \ .
\eea 
Thus, the quantity $\gamma$ measures how the proton net
production rate increases when the neutron fraction 
increases at fixed baryon density, \ie, $\gamma$ 
measures how fast the system reacts to a change in the 
chemical composition of matter. The quantity $\gamma^{-1}$ 
has a dimension of time and can be interpreted as a 
relaxation time of the system to its beta-equilibrium state.

The bulk viscosity in Equation~\eqref{eq:zeta} has the 
classic resonant form and for density oscillations of a
given frequency, $\omega$ attains a resonant maximum at 
the temperature where the relaxation rate matches the oscillation
frequency, $\gamma(T)=\omega$.

The value of the bulk viscosity at that maximum is
\bea\label{eq:zeta_max}
\zeta_{\rm max}=\frac{C^2}{2A\omega} \ ,
\eea 
which is independent of the microscopic interaction rates. 

In the regime of slow equilibration, where $\gamma\ll\omega$, 
the bulk viscosity takes the form
\bea\label{eq:zeta_slow}
\zeta_{\rm slow} = \frac{C^2}{A}
\frac{\gamma}{\omega^2} {= \dfrac{2 \gamma}{\omega}\, \zeta_{\rm max}}\, .
\eea 
In the fast equilibration regime $\gamma\gg\omega$,
and the bulk viscosity, in this case, reduces to
\bea\label{eq:zeta_fast}
\zeta_{\rm fast} = \frac{C^2}{A\gamma}
{= \dfrac{2\omega}{\gamma} \zeta_{\rm max} }\, .
\eea
The physical reason for this resonant maximum is easy to understand. 
In the limit where the relaxation rate is much smaller than the oscillation frequency
there would effectively be an additional conserved quantity since both proton number and neutron number would be conserved. The proton fraction would be independent of density, so density oscillations would not drive the system out of chemical equilibrium. There would then be no bulk viscosity.

In the opposite limit of fast equilibration or slow density
oscillations, $\gamma\gg\omega$, weak interactions are able
to restore the chemical equilibrium of matter on timescales much
smaller than the oscillation period. This means that the matter is
practically always beta-equilibrated 
while undergoing compression and rarefaction and,
therefore, will not experience any bulk viscosity.

Above we computed the bulk viscous pressure using its standard definition 
\bea\label{eq:Pi_inf}
\Pi =\delta P -\delta P_{\rm eq}
=\delta P(\lambda) -\delta P(\lambda\to\infty),
\eea 
where $\delta P$ is the shift of the pressure from its
static equilibrium value $P_0=P(n_{j0})$ for arbitrary
perturbations $\delta n_{j}$, and $\delta P_{\rm eq}$
is the instantaneous shift of the equilibrium pressure,
which depends on $\delta n_{j}^{\rm eq}$ and does not 
contribute to the bulk viscous pressure.
  
According to the two limiting cases of vanishing bulk 
viscosity discussed above the bulk viscous pressure 
can be defined also in an alternative way
\bea\label{eq:Pi_0}
\Pi =\delta P(\lambda) -\delta P(\lambda\to 0)
=\delta P -\delta P_{0},
\eea 
where $\delta P_{0}$ is the instantaneous shift of the pressure 
in a certain non-equilibrium state with conserved particle fractions 
which corresponds to the limit $\lambda\to 0$.
According to Equation~\eqref{eq:Pi_0} we can take as alternatives 
of the beta-equilibrium shifts $\delta n^{\rm eq}_j(t)$ the shifts $\delta n^0_j(t)$ given by Equation~\eqref{eq:cont_j}, as these are the solutions 
of exact balance equations~\eqref{eq:cont_n_weak} in the limit 
$\lambda\to 0$. This was just the choice of the equilibrium 
state in Ref.~\cite{Huang2010}. Then instead of 
Equation~\eqref{eq:delta_j'} we will have
\bea\label{eq:delta_j'_alt}
\delta n'_n(t) = \delta n'_\nu(t)=
-\delta n'_p(t) = -\delta n'_e(t)=
\frac{C}{A(i\omega+\gamma)}\frac{\gamma\theta}{i\omega}.
\eea 
Computing the bulk viscous pressure from Equation~\eqref{eq:Pi}
we obtain the same result for $\zeta$, as expected. 

\subsection{Urca process rates}
\label{sec:urca_rates}

As it was shown in Ref.~\cite{Alford2019b} the equilibration rate
$\lambda$ of the neutrino-trapped matter is dominated by the electron 
capture process because the neutron decay rate is exponentially 
damped at low temperatures as $\sim\exp(-\mu{_\nu}/T)$.
The microscopic $\beta$-equilibration rate of the electron 
capture process is given by~\cite{Alford2019b}
\bea \label{eq:Gammap_def}
\Gamma_{n}(\mu_\Delta) = 2\tilde G^2 
\prod_j \int\!\! \frac{d^3p_j}{(2\pi)^3}
\bar{f}(p_1){f}(p_2){f}(p_3)\bar{f}(p_4)\, 
(2\pi)^4\delta(p_1-p_2-p_3+p_4),
\eea 
where $\tilde{G}^2\equiv G_F^2\cos^2 \theta_c (1+3g_A^2)$, 
$G_F=1.166\cdot 10 ^{-5}$ GeV$^{-2}$ is the Fermi coupling 
constant, $\theta_c$ is the Cabibbo angle with 
$\cos\theta_c=0.974$, and $g_A=1.26$ is the axial-vector 
coupling constant; the index $j$ runs over the four
participating particles $j=\{n,p,e,\nu\}$, $f(p_j)$ are the 
Fermi distributions of particles and $\bar{f}(p_j)=1-f(p_j)$.
The inverse process $\Gamma_p$ will be given by an analogous 
expression by replacing all $f_j$ functions with $\bar{f_j}$.

For small departures from $\beta$-equilibrium $\mu_\Delta 
\ll T$ the imbalance between the direct and inverse 
rates can be linearized in $\mu_\Delta$ with the 
coefficient of the linear expansion given by~\cite{Alford2019b}
\bea\label{eq:lambda_final}
\lambda = \frac{ m^{*2}\tilde G^2 T^5}{8\pi^5}
\!\int_{-\infty}^{\infty} dy~g(y)\! 
\int_{z_0}^{\infty}  dz \, {\cal L}(y,z)
\int_{x_0}^{\infty}  dx\, (x+\alpha_{\nu})
(y+\alpha_e+x)f(x)\bar f(x+y),
\eea 
where $m^\ast$ is the effective nucleon mass, $\alpha_j = \mu_j^*/T$, 
$\mu_j^*$ are the effective chemical potentials of particles,  
$f(x)=(e^x+1)^{-1}$ and $g(x)=(e^x-1)^{-1}$ are the Fermi 
and the Bose distributions, respectively,
$x_0=(z-y-\alpha_e-\alpha_{\nu})/2$,
$z_0=|y+\alpha_e-\alpha_{\nu}|$, and 
\bea\label{eq:log_y_0}
 {\cal L}(y,z) = \ln \Bigg\vert \frac{1+\exp(-y_0)} 
 {1+\exp\left(-y_0-y\right)}\Bigg\vert,
\quad 
y_0 = \frac{m^*}{2Tz^2}\bigg(\alpha_n-\alpha_p+y
-z^2\frac{T}{2m^*}\bigg)^2-\alpha_p.
\eea
 
In the case of neutrino-transparent matter the equilibration rate 
$\lambda$ is given by a similar expression, see Refs.~\cite{Alford2019a,Alford2019b}
for details. The low-temperature 
limit of $\lambda$ in neutrino-trapped matter is given by~\cite{Alford2019b}
\bea \label{eq:lambda_deg_trap}
\lambda = \frac{1}{12\pi^3}m^{*2}\tilde G^2 T^2
\, p_{Fe}p_{F\nu} (p_{Fp}+p_{Fe}+p_{F\nu}-p_{Fn}),
\eea 
and in the neutrino-transparent matter~\cite{Haensel:2000vz,Alford2019a,Alford2019b}
\bea \label{eq:lambda_deg_trans}
\lambda = \frac{17}{240\pi} m^{*2} \tilde{G}^2 
T^4 \, p_{Fe}\,\theta( p_{Fp}+p_{Fe} -p_{Fn}),
\eea 
where $p_{Fj}$ are the Fermi-momenta of the particles.
The $\theta$-function in Equation~\eqref{eq:lambda_deg_trans}
blocks the direct Urca processes at low densities 
where the proton and electron Fermi momenta are not 
sufficiently large to guarantee the momentum conservation.
In the case where neutrinos are trapped in matter the 
momentum conservation can be always satisfied for
certain particle momenta and the rate is always finite.

\section{Numerical results}
\label{sec:results}

We start with the thermodynamics of nuclear matter, which is derived
from the relativistic density functional theory based on
phenomenological baryon-meson Lagrangians, for reviews
see~\cite{Weber_book,Sedrakian2007PrPNP}. We use the density-dependent
baryon-meson coupling model DDME2~\cite{Lalazissis2005} as applied to
finite-temperature nucleonic matter, see for details
Ref.~\cite{Colucci2013,Li_PLB_2018,Li2019ApJ}.

\subsection{Thermodynamics of nuclear matter}
\label{sec:DFT}

The Lagrangian density of the model considered reads
\bea\label{eq:lagrangian} 
{\cal L} & = &
\sum_N\bar\psi_N\bigg[\gamma^\mu \left(i\partial_\mu-g_{\omega}
\omega_\mu - \half g_{\rho}\vectau\cdot\vecrho_\mu\right)
- (m_N - g_{\sigma}\sigma)\bigg]\psi_N 
+ \sum_{\lambda}\bar\psi_\lambda(i\gamma^\mu\partial_\mu -
m_\lambda)\psi_\lambda\nonumber\\
 & + & \half \partial^\mu\sigma\partial_\mu\sigma-\half
m_\sigma^2\sigma^2 - \frac{1}{4}\omega^{\mu\nu}\omega_{\mu\nu} + \half
m_\omega^2\omega^\mu\omega_\mu -
\frac{1}{4}\vecrho^{\mu\nu}\vecrho_{\mu\nu} + \half
m_\rho^2\vecrho^\mu\cdot\vecrho_\mu, 
\eea 
where $N$ sums over nucleons, $\lambda$ -- over the leptons,
and $\psi_i$ are the fermionic Dirac fields with masses $m_i$. 
The meson fields $\sigma,\omega_\mu$, and $\vecrho_\mu$ are the
effective mediators of strong interaction between baryons,
$\omega_{\mu\nu}=\partial_\mu\omega_\nu-\partial_\nu\omega_\mu$ 
and $\vecrho_{\mu\nu}=\partial_\mu\vecrho_\nu-\partial_\nu\vecrho_\mu$ 
are the field strength tensors of $\omega_\mu$ and $\vecrho_\mu$ 
mesons, respectively, $m_{\sigma}$, $m_{\omega}$, and $m_{\rho}$ are 
meson masses, and $g_{i}$ are the baryon-meson coupling constants 
with $i=\sigma,\omega,\rho$.

\begin{figure}[t] 
\begin{center}
\includegraphics[width=0.45\columnwidth, keepaspectratio]{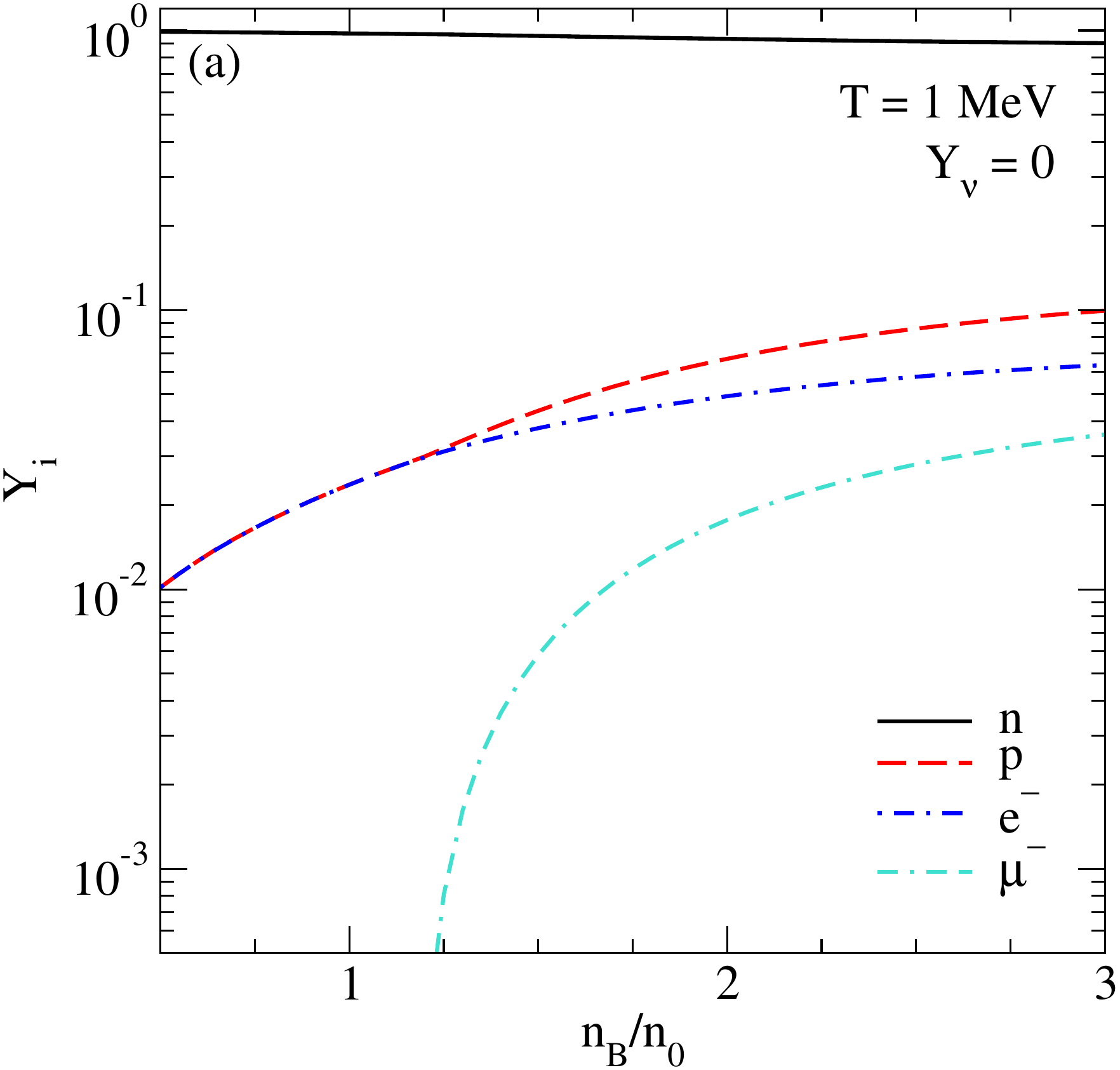}
\hskip 0.5cm
\includegraphics[width=0.45\columnwidth, keepaspectratio]{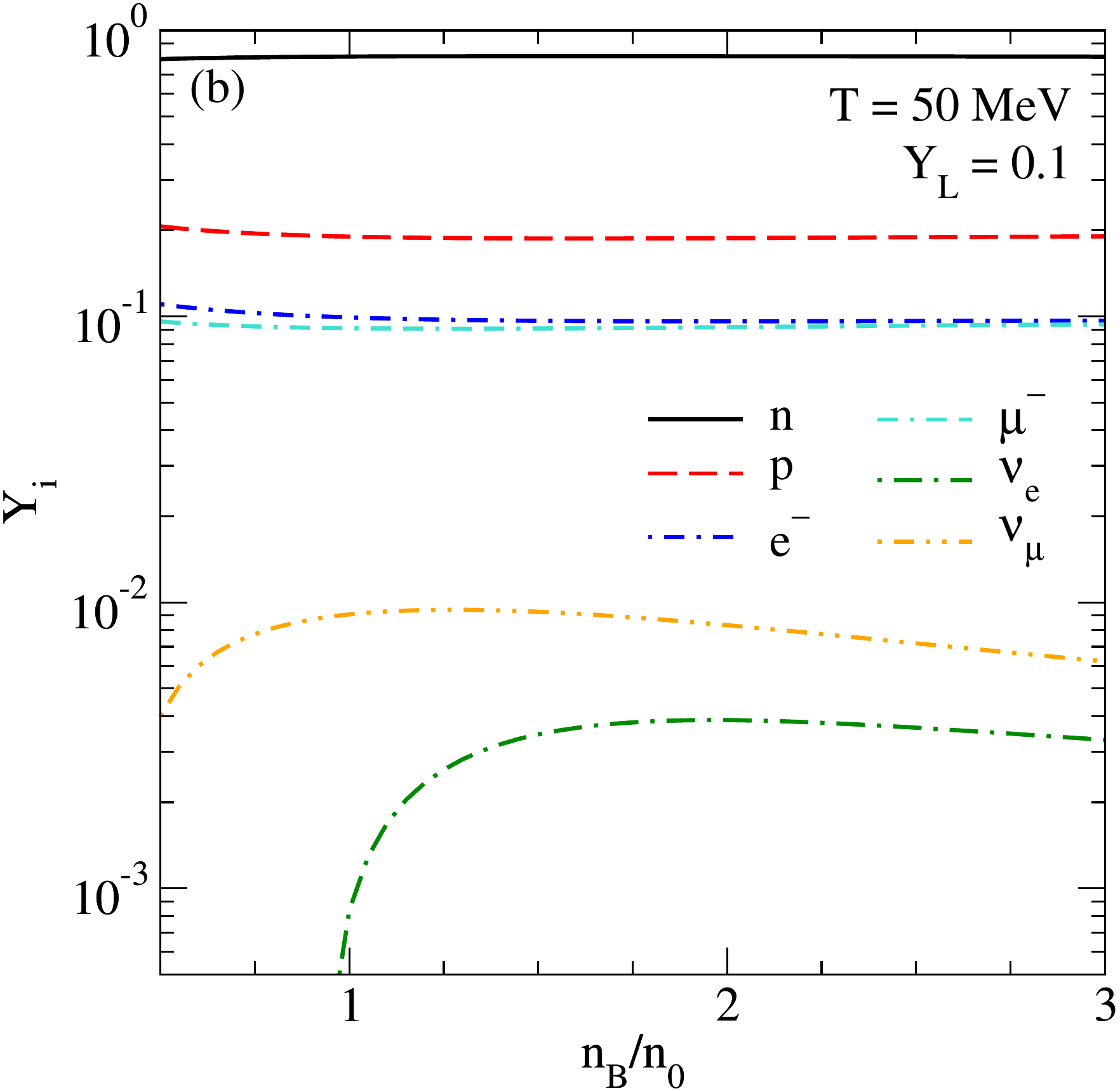}
\caption{ Particle fractions as functions of the baryon
density for (a) neutrino-transparent matter ($Y_\nu=0$) at
$T=1$\,MeV; (b) neutrino-trapped matter at $Y_L=0.1$ and $T=50$\,MeV. }
\label{fig:fractions} 
\end{center}
\end{figure}
\begin{figure}[!] 
\begin{center}
\includegraphics[width=0.45\columnwidth,keepaspectratio]{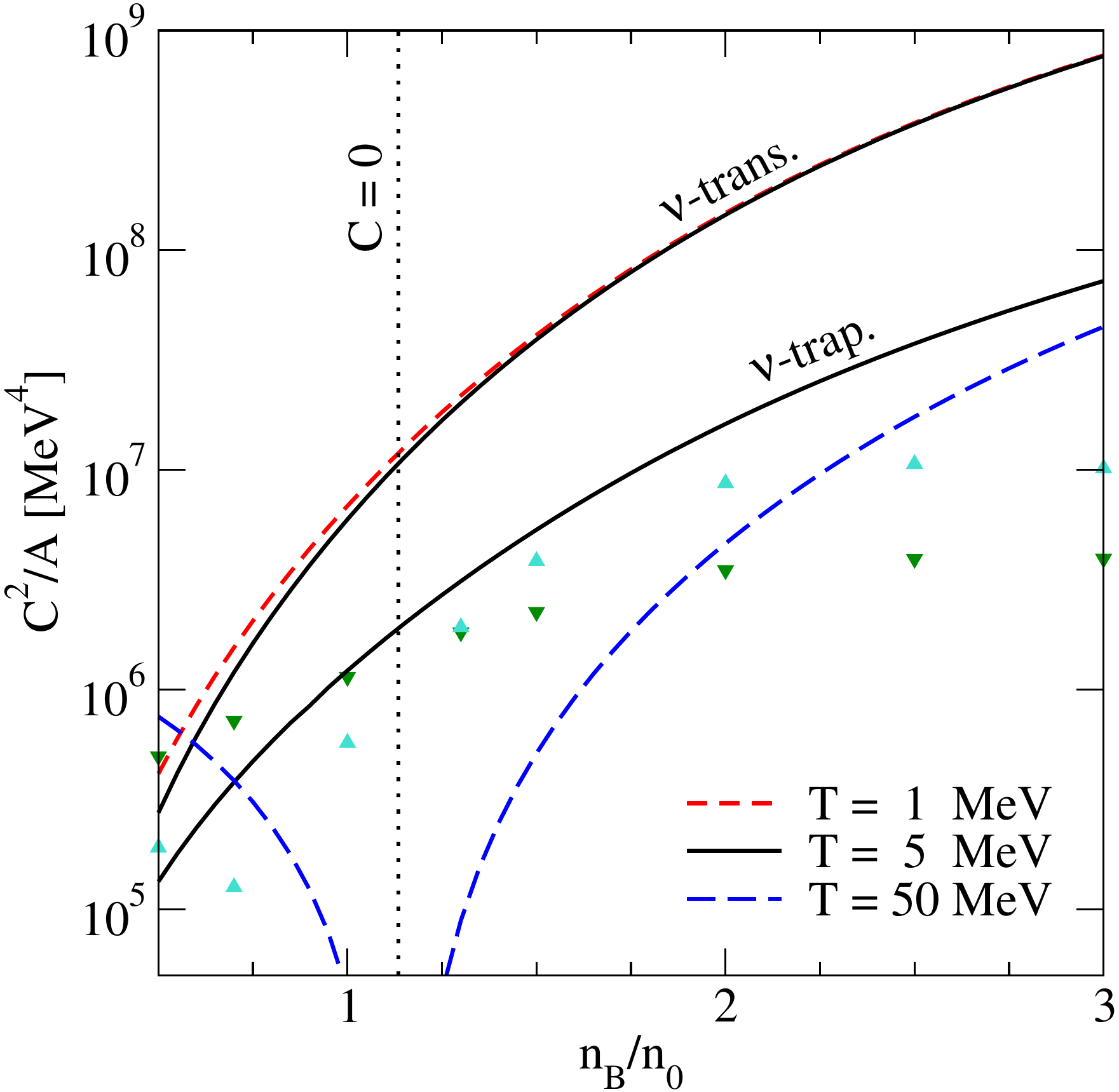}
\caption{ The susceptibility prefactor $C^2/A$ as a function
  of the baryon density for three values of the temperature
  for the neutrino-transparent matter (upper curves labelled as $\nu$-trans.) and neutrino-trapped matter (lower curves labelled as $\nu$-trap.). 
  The upper and lower triangles show the results of Ref.~\cite{Alford2019a} for models IUFSU and DD2, respectively.}
\label{fig:C2A_dens} 
\end{center}
\end{figure}
The density dependence of the particle fractions $Y_j = n_j/n_B$ is
shown in Fig.~\ref{fig:fractions}. The left panel is for a low
temperature $T=1$\,MeV where the system is neutrino-transparent,
and the right panel shows the results for neutrino-trapped matter at
temperature $T=50$\,MeV for lepton fraction fixed at $Y_L=0.1$.  In the
first case muons appear only above a certain baryon density
$n_B\gtrsim n_0$, where the condition $\mu_e\geq m_\mu\simeq 106$\,MeV
is satisfied, whereas in the neutrino-trapped case the muons' threshold
disappears.

Within the framework of the model above the susceptibilities $A$ and $C$ 
in the non-relativistic limit for nucleons are given by (see Ref.~\cite{Alford2019b} for details)
\bea\label{eq:A_fianl}
A &=& \frac{1}{{I}_{2n}}+\frac{1}{{I}_{2p}}+
\frac{1}{{I}_{2e}}+\frac{2}{{I}_{2\nu}}
+\left(\frac{g_\rho}{m_\rho}\right)^2,\\
\label{eq:C_fianl}
C &=& \frac{n_n}{{I}_{2n}}-\frac{n_p}{{I}_{2p}}-\frac{n_e}{{I}_{2e}}
+2\frac{n_\nu}{{I}_{2\nu}}-g_\rho \rho_{03}
+\frac{g_\sigma \sigma}{2m^{*2}}
\left(\frac{{I}_{4n}}{{I}_{2n}} - \frac{{I}_{4p}}{{I}_{2p}}\right),
\eea
where 
\bea\label{eq:I_nonrel}
{I}_{q i}= \frac{1}{\pi^2 T}\int_0^\infty p^q dp \,f_i \bar f_i,
\eea 
$\sigma$ is the $\sigma$ meson mean field and $\rho_{03}$ 
is the $\rho$ meson mean field which is non-zero
in asymmetric nuclear matter. These expressions for susceptibilities 
are derived for isothermal density perturbations, which is the case
only if the thermal conduction is fast enough to smoothen the 
temperature gradients during one period of
oscillation~\cite{Alford2018a,Alford2019b}.  
This might happen, \eg, in the presence of turbulent flows in the merger region which could generate temperature and density variations on distance scales
of the order of a few hundred meters.
An order-of-magnitude estimate of 
thermal relaxation timescale is given
in Ref.~\cite{Alford2018a}, which is
1 sec $\times (z_{\rm typ}/{\rm km})^2 (T/10\, {\rm MeV})^2$, where
$z_{\rm typ}$ is the typical length 
scale of thermal gradients. Assuming
$z_{\rm typ}\simeq 100$ m and temperatures $1\div 10$ MeV (this is the temperature range where the bulk viscosity is relevant to mergers, see below), the thermal relaxation time will lie in the interval 0.1-10~ms, which is below the characteristic timescale of binary neutron star mergers. Thus, for thermal gradients on this distance scale the assumption of isothermal matter is the relevant one. 
On the scales over which thermal conduction is inefficient the 
matter should be treated as iso-entropic. The isothermal and 
adiabatic susceptibilities differ at most by a factor of 2 in 
the relevant density and temperature range, see 
Ref.~\cite{Alford2019a} for further details. 

Figure~\ref{fig:C2A_dens} shows the ratio $C^2/A$ of 
susceptibilities as a function of density for three values of 
the temperature.  The susceptibility $C$ is an increasing function 
of density. At sufficiently high temperatures $T\gtrsim 30$\,MeV, 
it is negative at low density and crosses zero
at a temperature-dependent critical density 
where the proton fraction reaches a minimum as a 
function of the density. At that point, the system becomes 
scale-invariant, as the compression does not drive the matter 
out of beta equilibrium, and the bulk viscosity vanishes at 
that critical point. 

The ratio $C^2/A$ grows rapidly with the density in both cases of
neutrino-transparent and neutrino-trapped matter and is sensitive to
the temperature only close to the point where $C=0$. We see that
$C^2/A$ is approximately an order of magnitude smaller in the
neutrino-trapped matter the reason being much larger values of $A$ 
dominated by the contribution of neutrinos as compared to neutrino-transparent matter.

\subsection{Beta relaxation rates and bulk viscosity}
\label{sec:bulk_visc}

\begin{figure}[t] 
\begin{center}
\includegraphics[width=0.435\columnwidth,keepaspectratio]{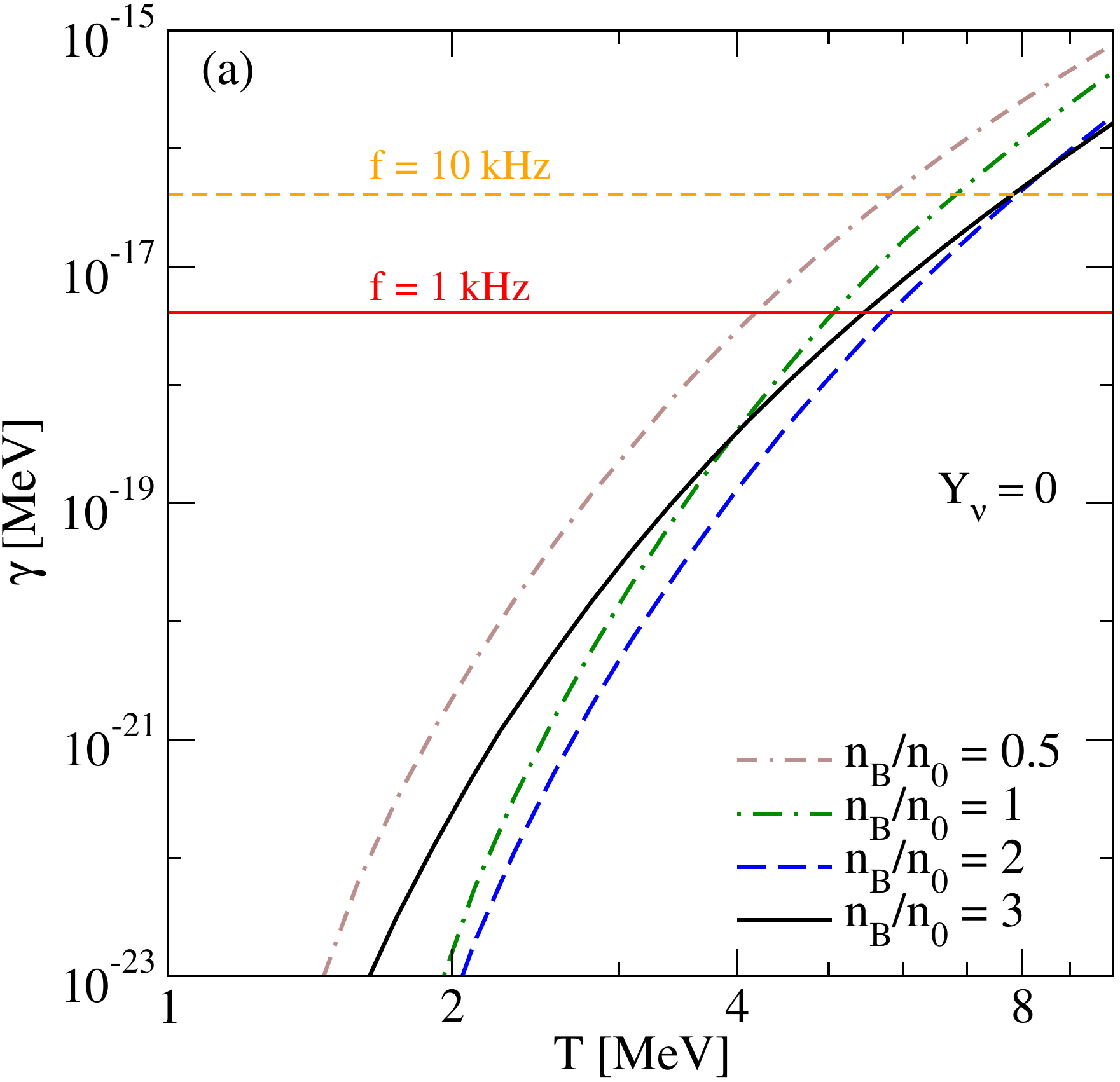}
\hskip 0.5cm
\includegraphics[width=0.45\columnwidth,keepaspectratio]{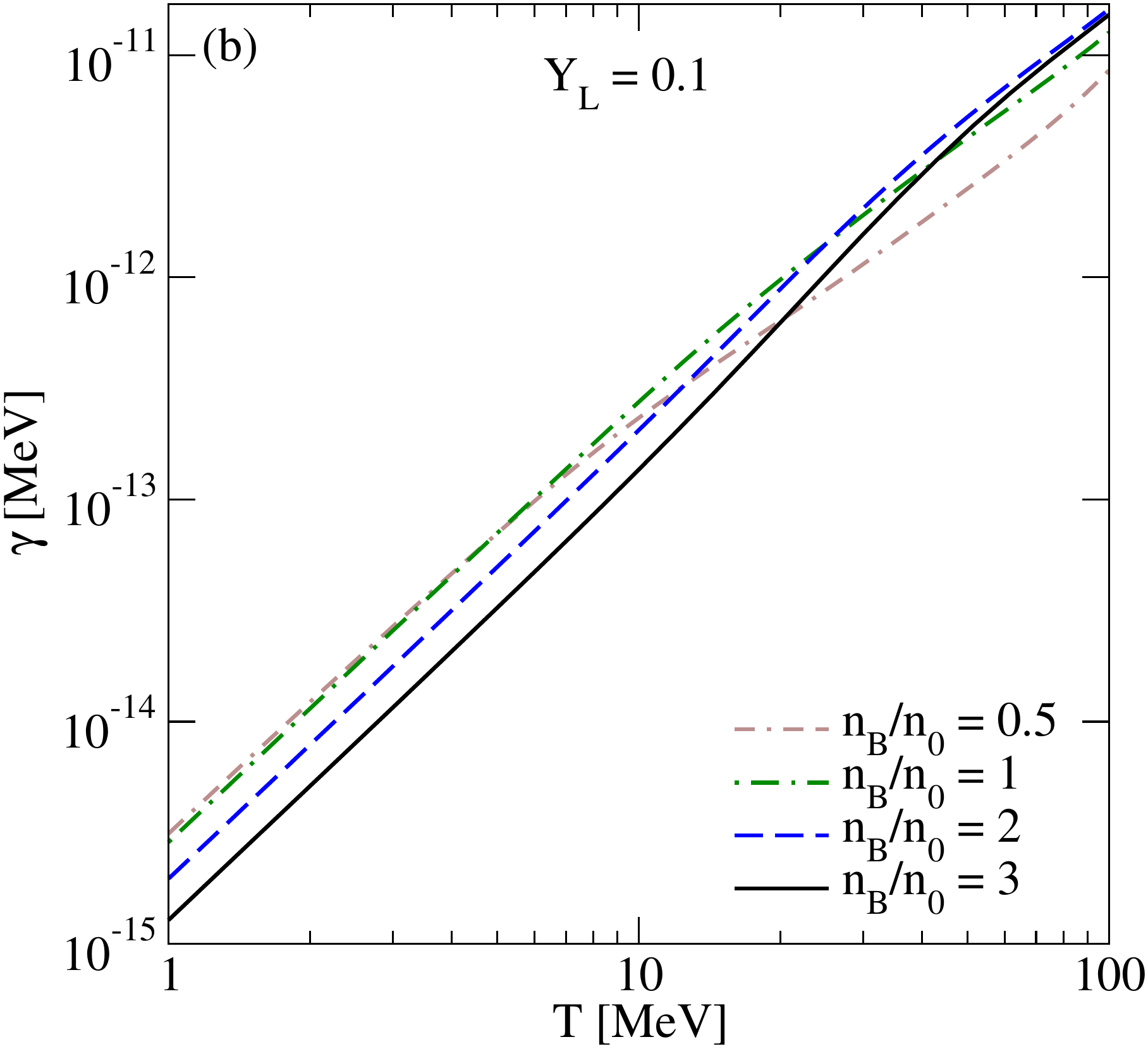}
\caption{ The beta-relaxation rate $\gamma$ as a function of the
  temperature for (a) neutrino-transparent and (b) neutrino-trapped 
  regimes for several values of baryon density. Note the very different $y$-axis scales. 
  The horizontal lines in panel (a) correspond to the fixed values of oscillation frequency 
  $f=1$ kHz (solid line) and $f=10$ kHz (dashed line).}
\label{fig:gamma_temp} 
\end{center}
\end{figure}

Figure~\ref{fig:gamma_temp} shows the relaxation rate $\gamma=\lambda A$
[Equations~\eqref{eq:lambda_def} and \eqref{eq:gamma}] as a
function of temperature for various densities. The equilibration rate 
$\lambda$ of the neutrino-trapped matter is dominated by the electron capture
process because the neutron decay rate is exponentially damped at low
temperatures as $\lambda_1\sim\exp(-\mu{_\nu}/T)$, whereas $\lambda_2$
has approximately a quadratic increase with the temperature as
suggested by Equation~\eqref{eq:lambda_deg_trap}. It is
seen from Fig.~\ref{fig:gamma_temp}(b) that the relaxation 
rate $\gamma$ of the neutrino-trapped matter is several
orders of magnitude larger than the oscillation frequencies
$f=\omega/2\pi\simeq 1$ kHz which are typical to neutron star
mergers. This means that the neutrino-trapped matter is always in the
fast equilibration regime where the bulk viscosity is independent of
the oscillation frequency and is given by Equation~\eqref{eq:zeta_fast}.

\begin{figure}[t] 
\begin{center}
\includegraphics[width=0.45\columnwidth,keepaspectratio]{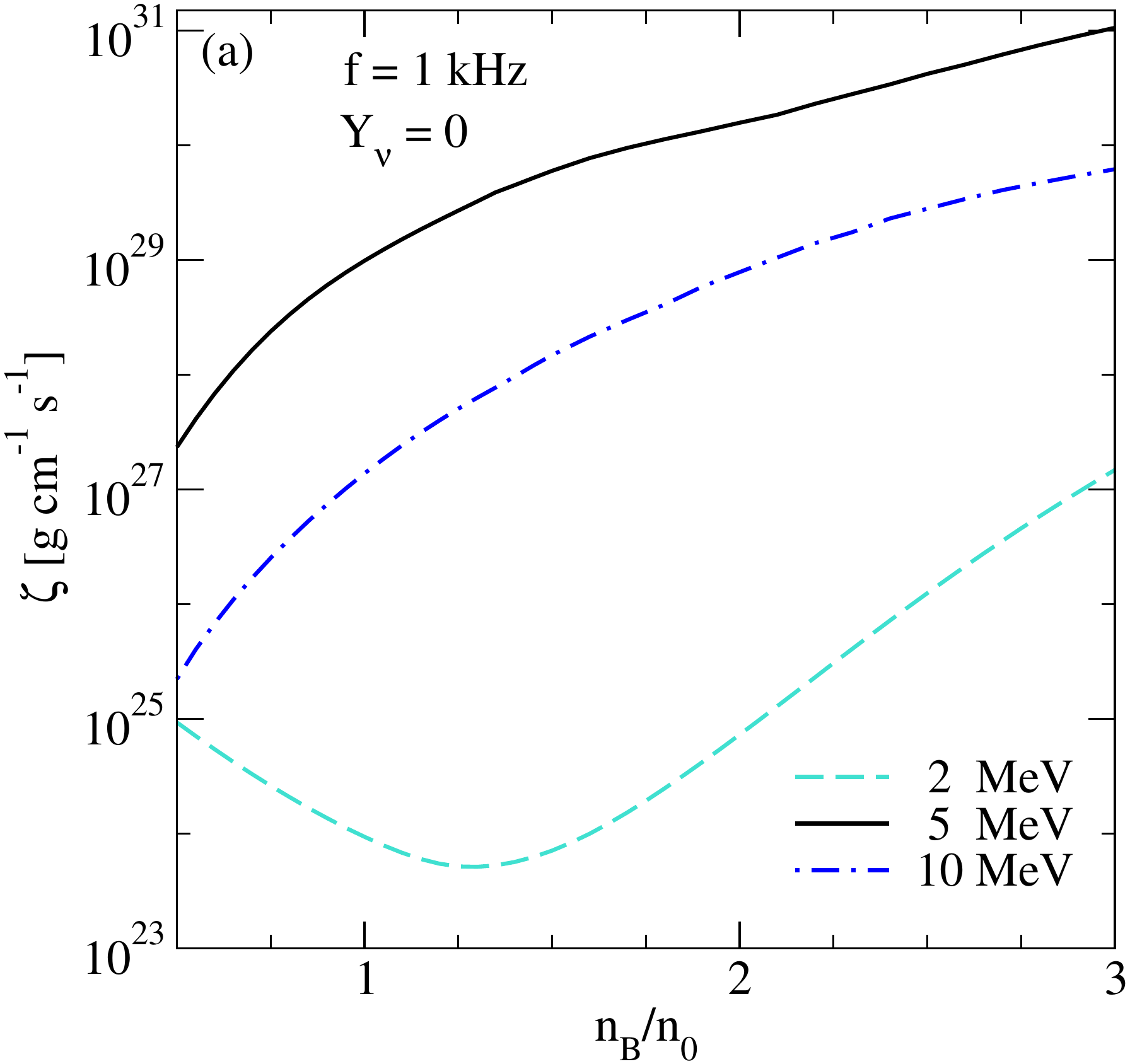}
\hskip 0.5cm
\includegraphics[width=0.45\columnwidth,keepaspectratio]{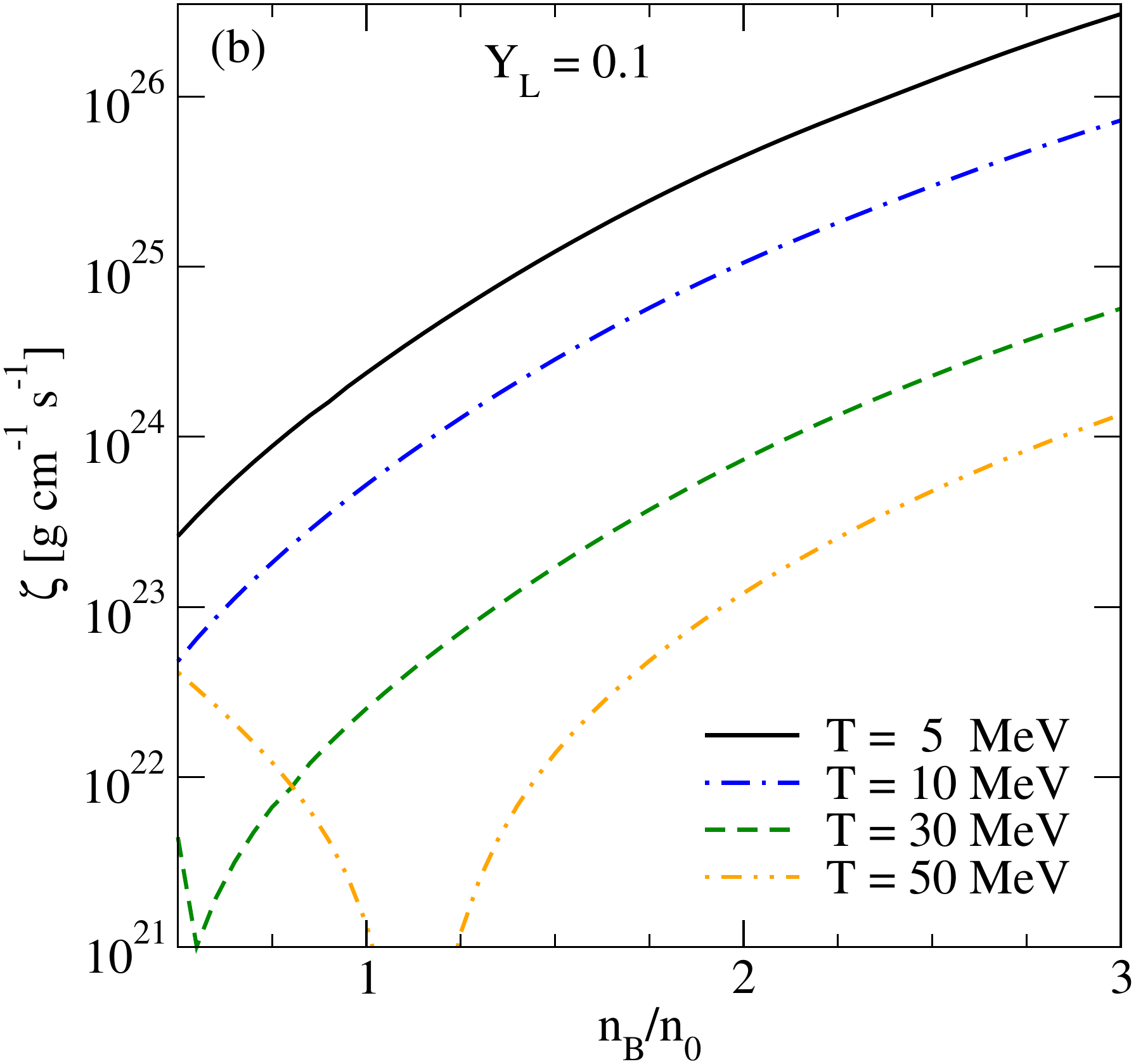}
\caption{ The density dependence of the bulk viscosity for 
  (a) neutrino-transparent matter; (b) neutrino-trapped matter at
  $Y_L=0.1$ for various values of the temperature. The oscillation frequency is $f=1$ kHz
  in panel (a); the bulk viscosity in panel (b) is frequency independent.
  Note the very different $y$-axis scales.}
\label{fig:zeta_dens} 
\end{center}
\end{figure}
\begin{figure}[!] 
\begin{center}
\includegraphics[width=0.45\columnwidth, keepaspectratio]{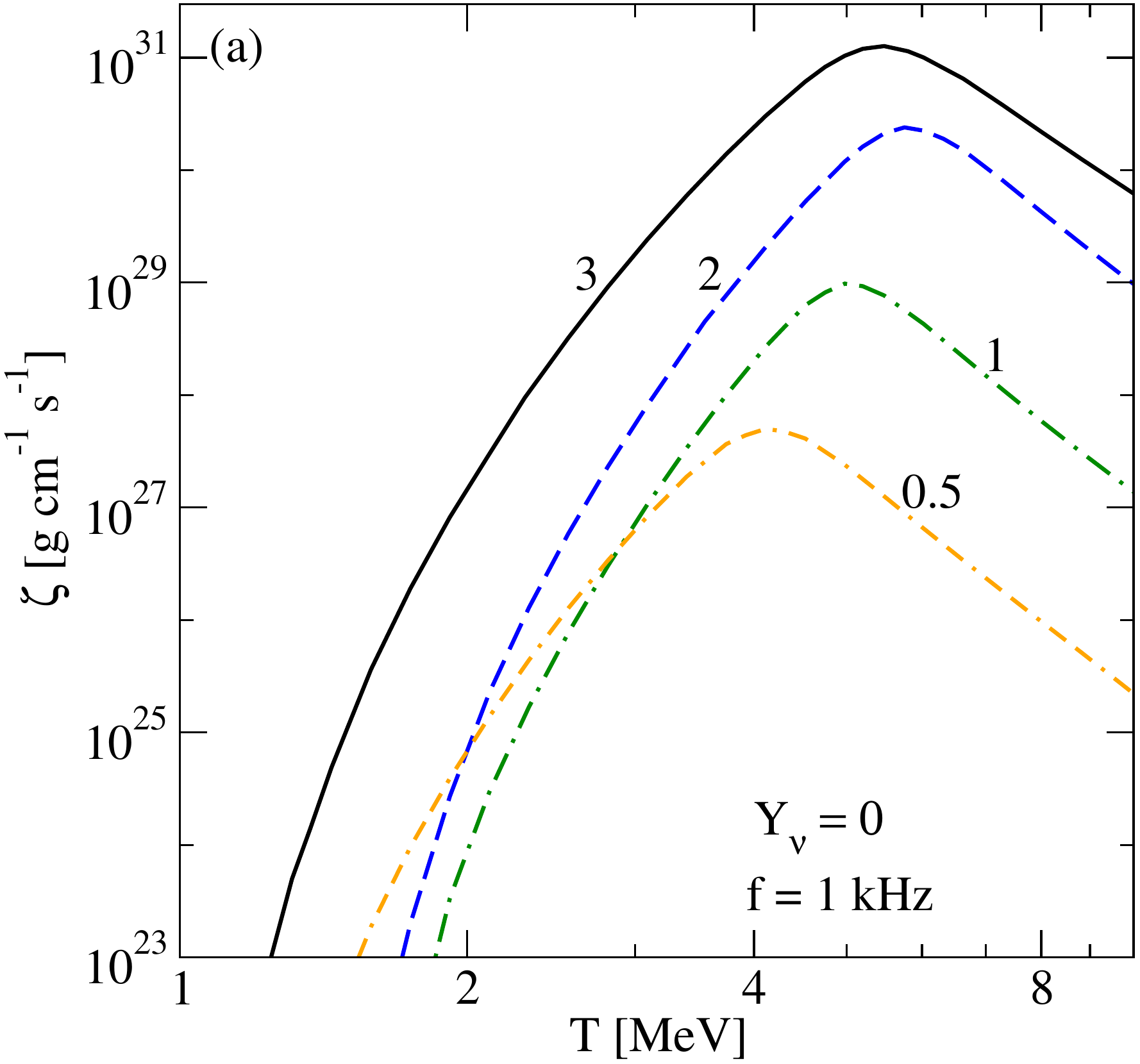}
\hskip 0.5cm
\includegraphics[width=0.465\columnwidth, keepaspectratio]{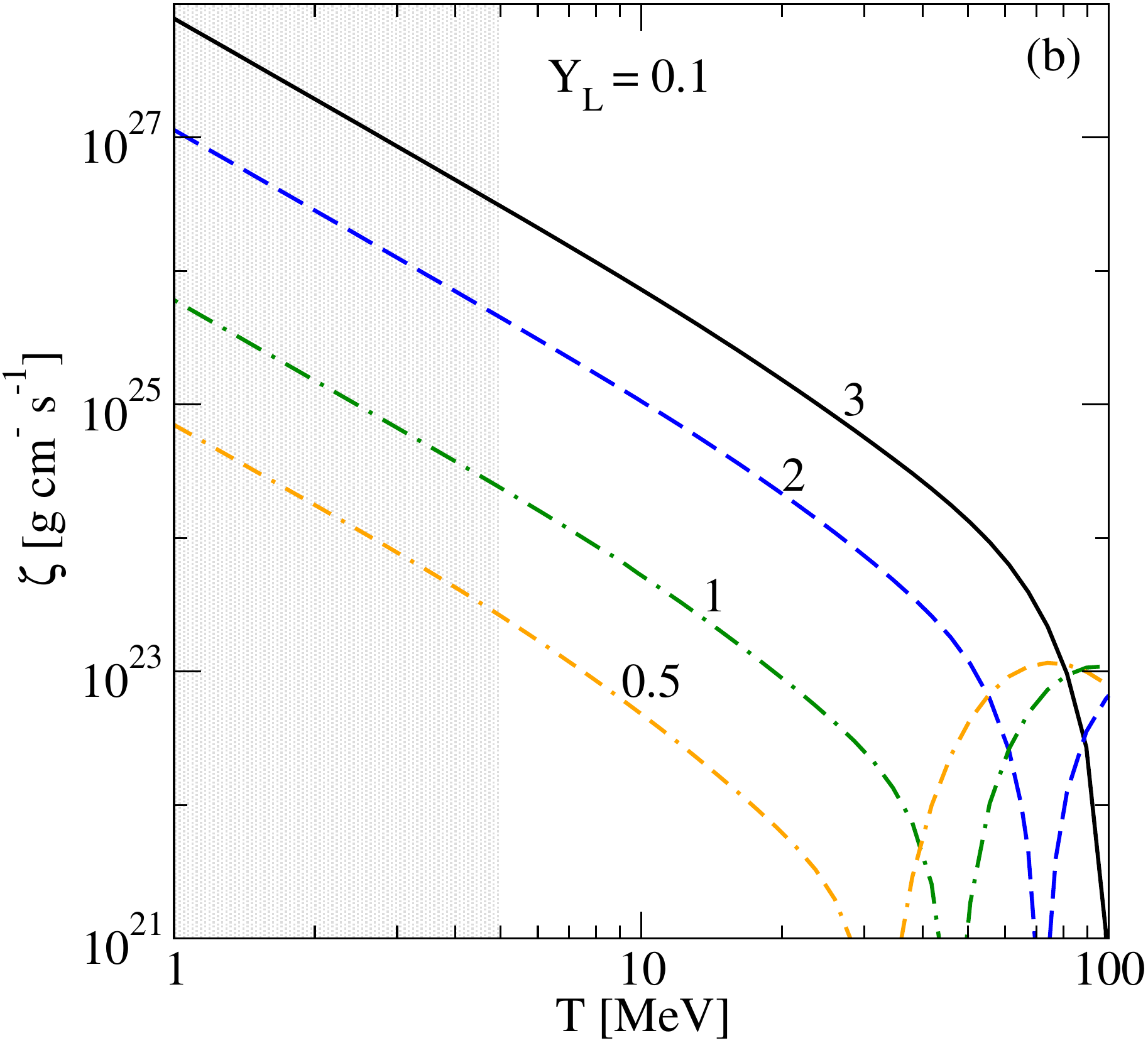}
\caption{ The temperature dependence of the bulk viscosity 
for several values of the baryon density marked on the plots 
for (a) neutrino-transparent matter; (b) neutrino-trapped matter 
at $Y_L=0.1$. The oscillation frequency is $f=1$ kHz in panel (a); the bulk viscosity
 in panel (b) is frequency independent.
Note the very different $y$-axis scales.  The shaded region in panel (b) 
indicates that the results therein are extrapolations to the regime where they are not applicable.
}
\label{fig:zeta_temp} 
\end{center}
\end{figure}

In the neutrino-transparent regime, in contrast, the relaxation rate 
is much slower, reaching values in the kHz range where it
can resonate with typical density oscillations in mergers.
We see this in Figure~\ref{fig:gamma_temp}(a) where the 
relaxation rate $\gamma$ crosses the $\omega = 2\pi$\,kHz (corresponding to $f=1$\,kHz) line 
at temperatures $4\div 5$\,MeV indicating that the neutrino-transparent
matter possesses a resonant maximum at those temperatures, as it was
found also in Refs.~\cite{Alford2018a,Alford2019a}.


\begin{figure}[t] 
\begin{center}
\includegraphics[width=0.45\columnwidth, keepaspectratio]{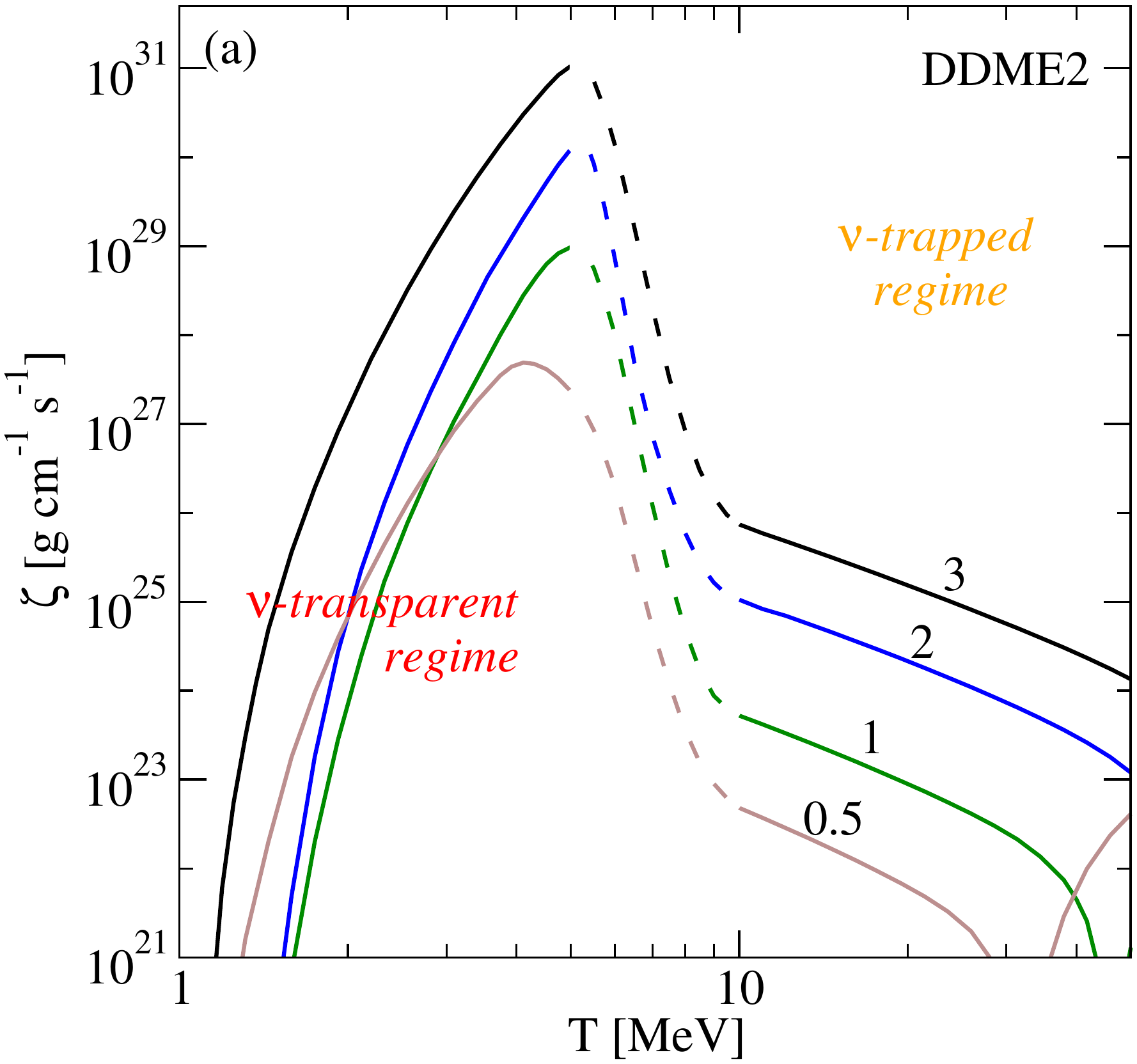}
\hskip 0.5 cm
\includegraphics[width=0.45\columnwidth, keepaspectratio]{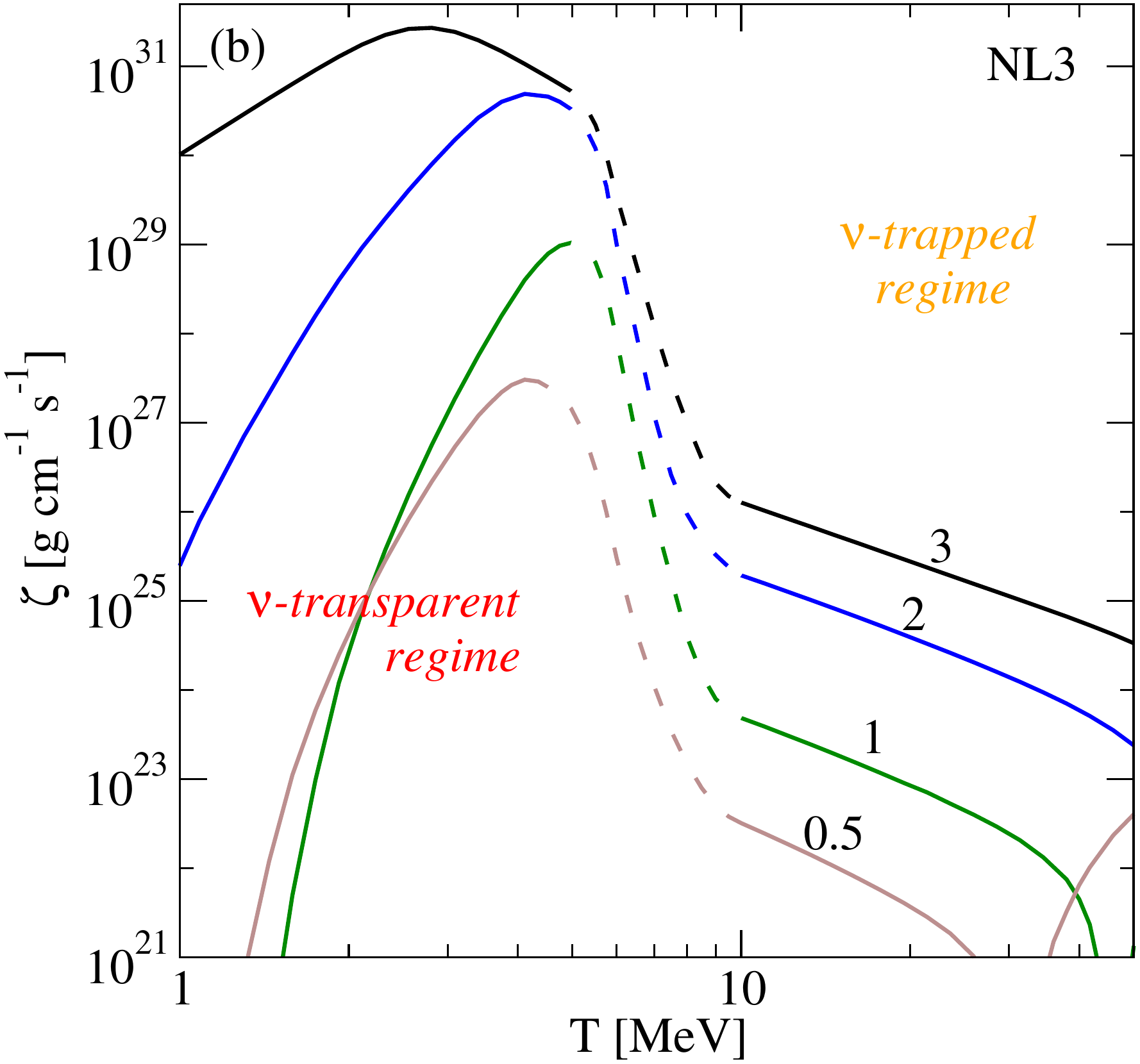}
\caption{ The bulk viscosity as a function of temperature for
 several values of the baryon density marked on the plots for
 (a) model DDME2; (b) model NL3. In the temperature range 
 $5\leq T\leq 10$\,MeV the bulk viscosity is obtained by 
 interpolating between the results of neutrino-transparent 
 and neutrino-trapped matter. The oscillation frequency is 
 fixed at $f=1$ kHz, and the lepton fraction is fixed at $Y_L=0.1$. }
\label{fig:zeta_temp_final} 
\end{center}
\end{figure}

The density dependence of the bulk viscosity is shown in 
Fig.~\ref{fig:zeta_dens}. The oscillation frequency is fixed at 
$f=1$ kHz in the case of neutrino-transparent matter whereas
the neutrino-trapped matter features a frequency-independent
bulk viscosity, as discussed above.
The bulk viscosity of neutrino-transparent matter mainly increases 
with the density, the increase being faster at low temperatures 
$T\lesssim 3$\,MeV where $\omega\gg \gamma$, as seen from 
Fig.~\ref{fig:gamma_temp}.

The density dependence of the bulk viscosity of neutrino-trapped matter mainly follows that of the susceptibility $C^2/A$ because $\gamma$ depends weakly on the baryon density in this case. At sufficiently high temperatures $T\gtrsim 30$\,MeV there are sharp drops of the bulk viscosity to zero related to the fact that the matter becomes scale-invariant at certain critical values of density. Note that for temperature range $5\div 10$\,MeV we show the results for both neutrino-transparent and neutrino-trapped cases to account for the uncertainty in the exact value of neutrino trapping temperature $T_{\rm tr}$ which is supposed to lie in that range.

Figure~\ref{fig:zeta_temp} plots the dependence of the bulk viscosity
on the temperature. As already discussed above, the bulk viscosity of
neutrino-transparent matter attains its maximum at the temperature
where {$\gamma(T)=\omega$}, whereas the temperature dependence of the bulk viscosity
in the neutrino-trapped matter is mainly decreasing (up to point where
the matter becomes scale-invariant) because the relaxation rate is 
already too fast: the resonant maximum would be at lower temperatures. 
Since the relaxation rate $\gamma$ rises as $T^2$ [Equation~\eqref{eq:lambda_deg_trap}] 
we expect [from  Equation~\eqref{eq:zeta_fast}] that
$\zeta\propto T^{-2}$ in this regime.

In Fig.~\ref{fig:zeta_temp_final}, panel (a) we combine the results
obtained for neutrino-transparent and neutrino-trapped matter by
interpolating the results between these two regimes in the temperature
range $5\leq T\leq 10$\,MeV which is regarded as the transition
region. Close to the transition temperature, the bulk viscosity is much 
larger in the neutrino transparent regime because of much lower beta
relaxation rate and larger susceptibility $C^2/A$ as well. As a
result, the resonant peak of the bulk viscosity occurs always at or
below the neutrino-trapping temperature. Hence we can already
anticipate that the bulk viscosity is going to play an important role
in the dynamics of neutron star mergers in the regime of
neutrino-transparent rather than neutrino-trapped matter.

For comparison, we show also the bulk viscosity of nuclear matter 
for an alternative  model NL3~\cite{Lalazissis1997} in panel (b) of
Fig.~\ref{fig:zeta_temp_final}. This model has density-independent 
meson-nucleon couplings but contains non-linear terms in the $\sigma$-meson 
field. The results obtained within the two models differ mainly
in the low-temperature regime, where the  model NL3 features much higher
viscosities than the model DDME2 above the saturation density. The reason for this is that NL3 
model has a threshold of direct Urca opening at around $n_B\simeq 2.5 n_0$ 
whereas the model DDME2 does not have a threshold up to densities 
$n_B=5n_0$. Because of the threshold, the model NL3 has much faster relaxation 
rates at $n_B\gtrsim 2n_0$ than the model DDME2, and, as
a consequence, the maximum of bulk viscosity is shifted to lower
values of the temperature as compared to the DDME2 model.

\subsection{Estimating damping timescales}
\label{sec:damping}

\begin{figure}[t] 
\begin{center}
\includegraphics[width=0.45\columnwidth, keepaspectratio]{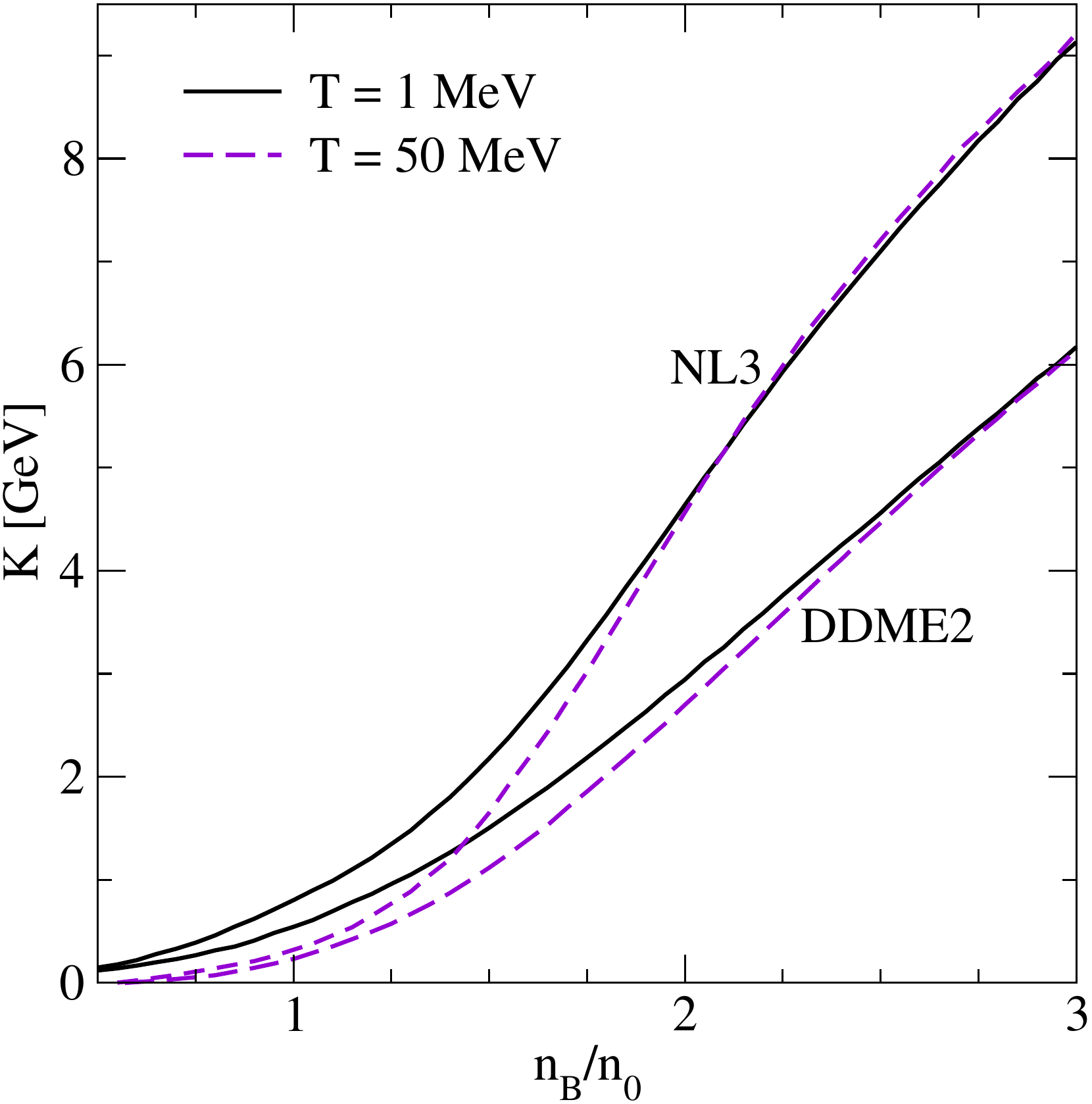}
\caption{ The isothermal incompressibility of nuclear matter as a function of density for several temperatures for models DDME2 and NL3. }
\label{fig:compress} 
\end{center}
\end{figure}

In this subsection we examine the characteristic timescales
of damping of density oscillations by the bulk viscosity in neutron 
star mergers. The energy density stored in baryonic oscillations 
with amplitude $\delta n_B$ is given by
\bea\label{eq:osc_energy}
\epsilon =\frac{1}{2} \frac{\partial^2\varepsilon}{\partial n_B^2}
(\delta n_B)^2=\frac{K}{18}\frac{(\delta n_B)^2}{n_B},
\eea 
where 
\bea\label{eq:compress}
K=9n_B\frac{\partial^2\varepsilon}{\partial n_B^2}
\eea 
is the isothermal incompressibility of nuclear matter. 
The energy dissipation rate by the bulk viscosity per unit volume is 
\bea\label{eq:energy_diss}
\frac{d\epsilon}{dt}=\frac{\omega^2\zeta}{2} 
\left(\frac{\delta n_B}{n_B}\right)^2.
\eea
The characteristic timescale required for damping of oscillations
$\tau_\zeta =\epsilon/(d\epsilon/dt)$ is then given by
\bea\label{eq:damping_time}
\tau_{\zeta} =\frac{1}{9}\frac{Kn_B}{\omega^2\zeta}.
\eea 

The incompressibility of nuclear matter is shown in Fig.~\ref{fig:compress} for the two parametrizations discussed above. It is an increasing function of the density and at low densities decreases from its value at zero temperature as the temperature is increased.

\begin{figure}[t] 
\begin{center}
\includegraphics[width=0.45\columnwidth, keepaspectratio]{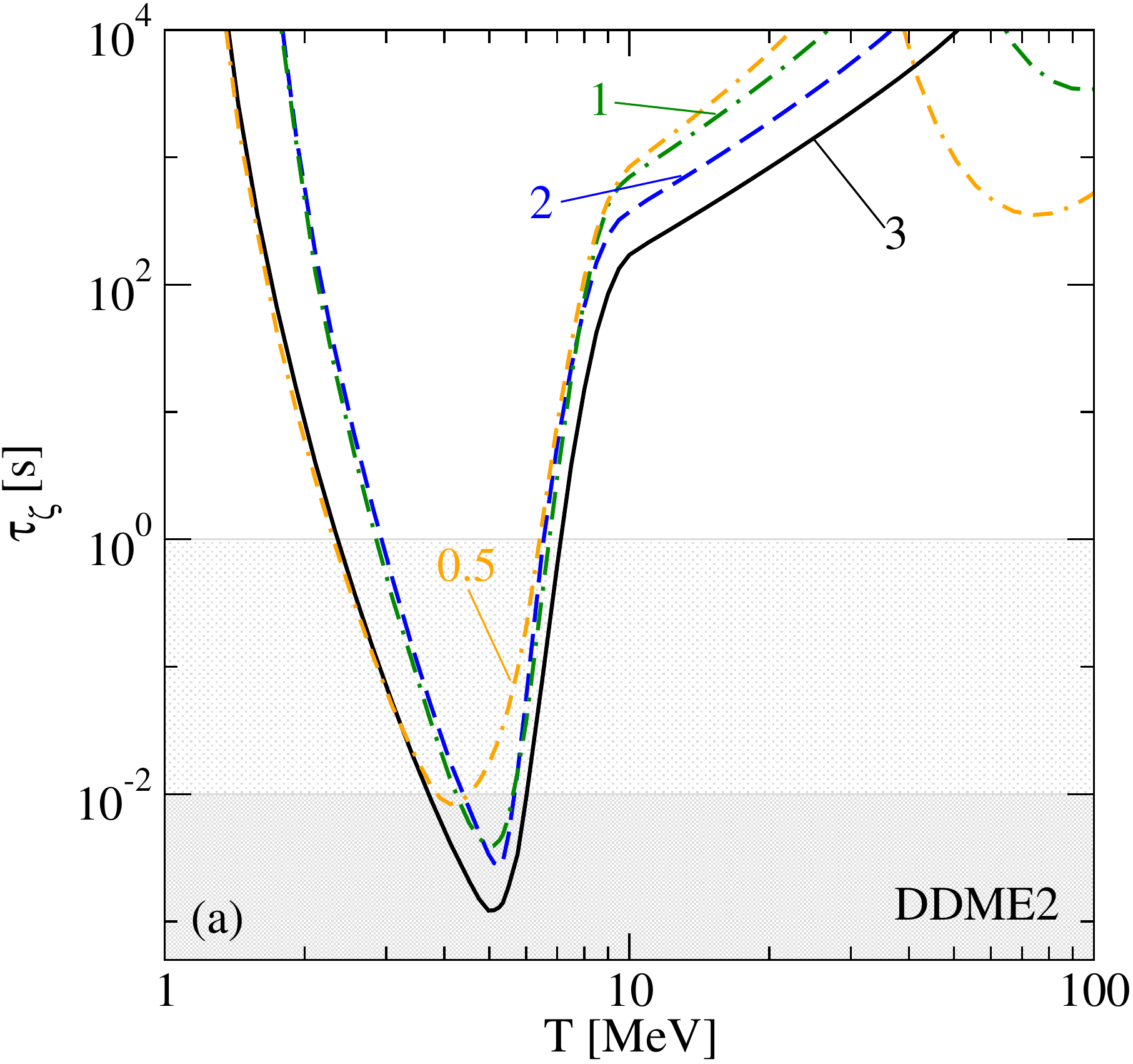}
\hskip 0.5cm
\includegraphics[width=0.45\columnwidth, keepaspectratio]{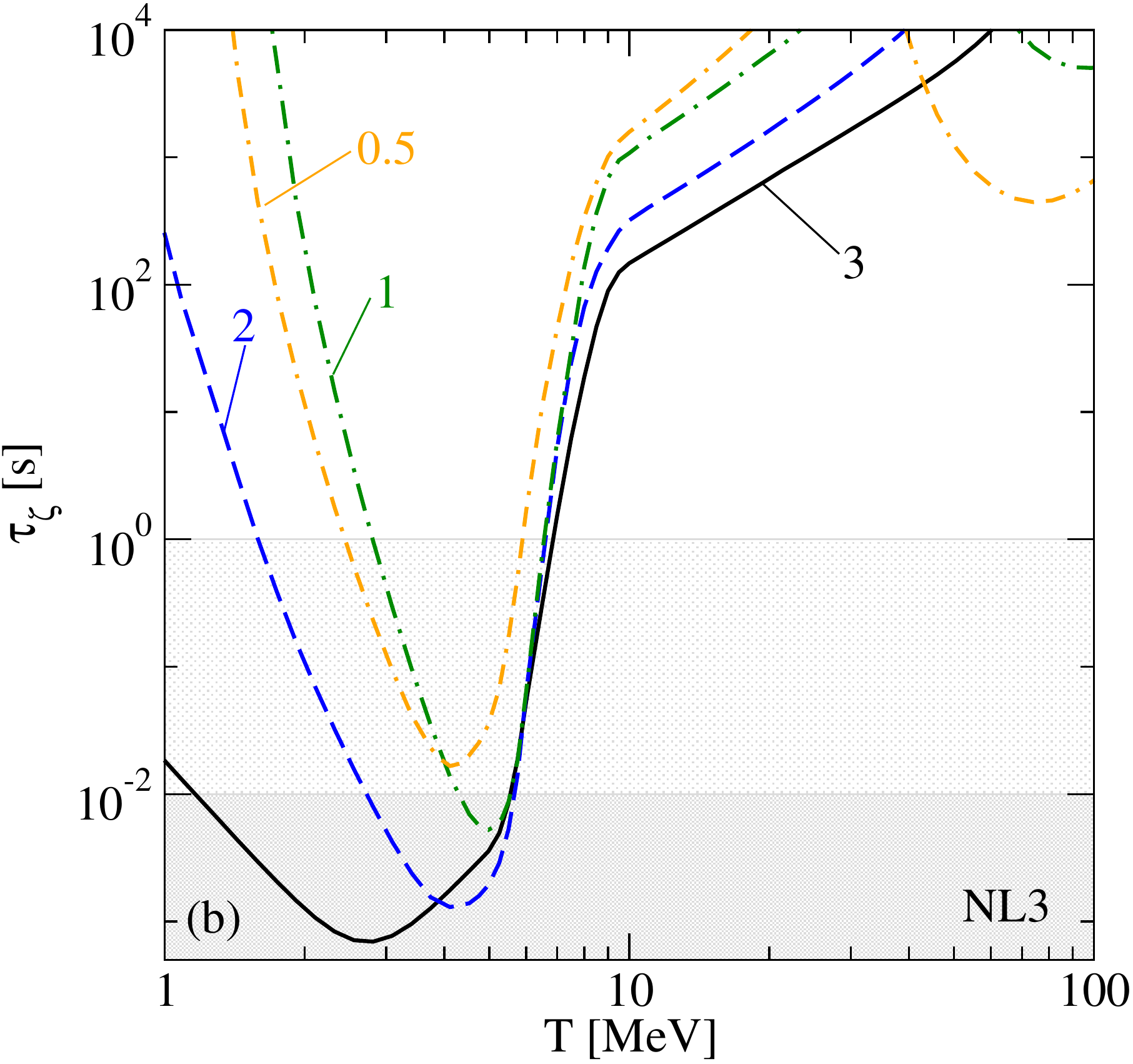}
\caption{ The oscillation damping timescale as a function of temperature 
for various densities and for frequency fixed at $f=1$ kHz. We interpolate between the neutrino-transparent and neutrino-trapped regime, as in Fig.~\ref{fig:zeta_temp_final}.
Panel (a): model DDME2; (b): model NL3. 
}
\label{fig:tau_damp1} 
\end{center}
\end{figure}

Figures~\ref{fig:tau_damp1}a and ~\ref{fig:tau_damp1}b show, for two EoSs, the damping timescales of oscillations with frequency $f=1$\,kHz. We use the interpolation of bulk viscosity between neutrino-transparent and neutrino-trapped regimes shown in Fig.~\ref{fig:zeta_temp_final}.  As nuclear
incompressibility depends weakly on the temperature, the temperature
dependence of the damping timescale closely follows that of the
bulk viscosity. As a result, the damping timescale attains a minimum
at the temperature where the bulk viscosity has a maximum for a fixed
value of the density. The minimal value of the damping timescale is  
\bea\label{eq:damping_time_min}
\tau_{\zeta}^{\rm min} =\frac{2}{9\omega}\frac{Kn_B}{C^2/A}.
\eea 
In the limits of slow and fast equilibration the damping 
timescale is given by 
\bea\label{eq:damping_time_limits}
\tau^{\rm slow}_\zeta = \frac{1}{9\gamma}\frac{Kn_B}{C^2/A},\qquad
\tau^{\rm fast}_\zeta =\frac{\gamma }{9\omega^2}
\frac{Kn_B}{C^2/A}.
\eea 
Thus, in contrast to the bulk viscosity, the damping timescale
becomes frequency-independent in the low-temperature (slow equilibration) 
regime, and decreases with the frequency at high temperature 
(fast equilibration) regime.

The density dependence of $\tau_{\zeta}$ reflects the density
dependence of the ratio of nuclear incompressibility and the 
bulk viscosity. The density dependence of these two quantities almost compensates each other in the neutrino-transparent regime of the DDME2 model. In the case of NL3 model the density dependence of the bulk viscosity dominates and the damping
timescale mainly decreases with density. The reason for this is the increase of the reaction rates with density as a result of fast opening of phase space for
direct Urca reactions in the NL3 model.

\begin{figure}[!] 
\begin{center}
\includegraphics[width=0.45\columnwidth, keepaspectratio]{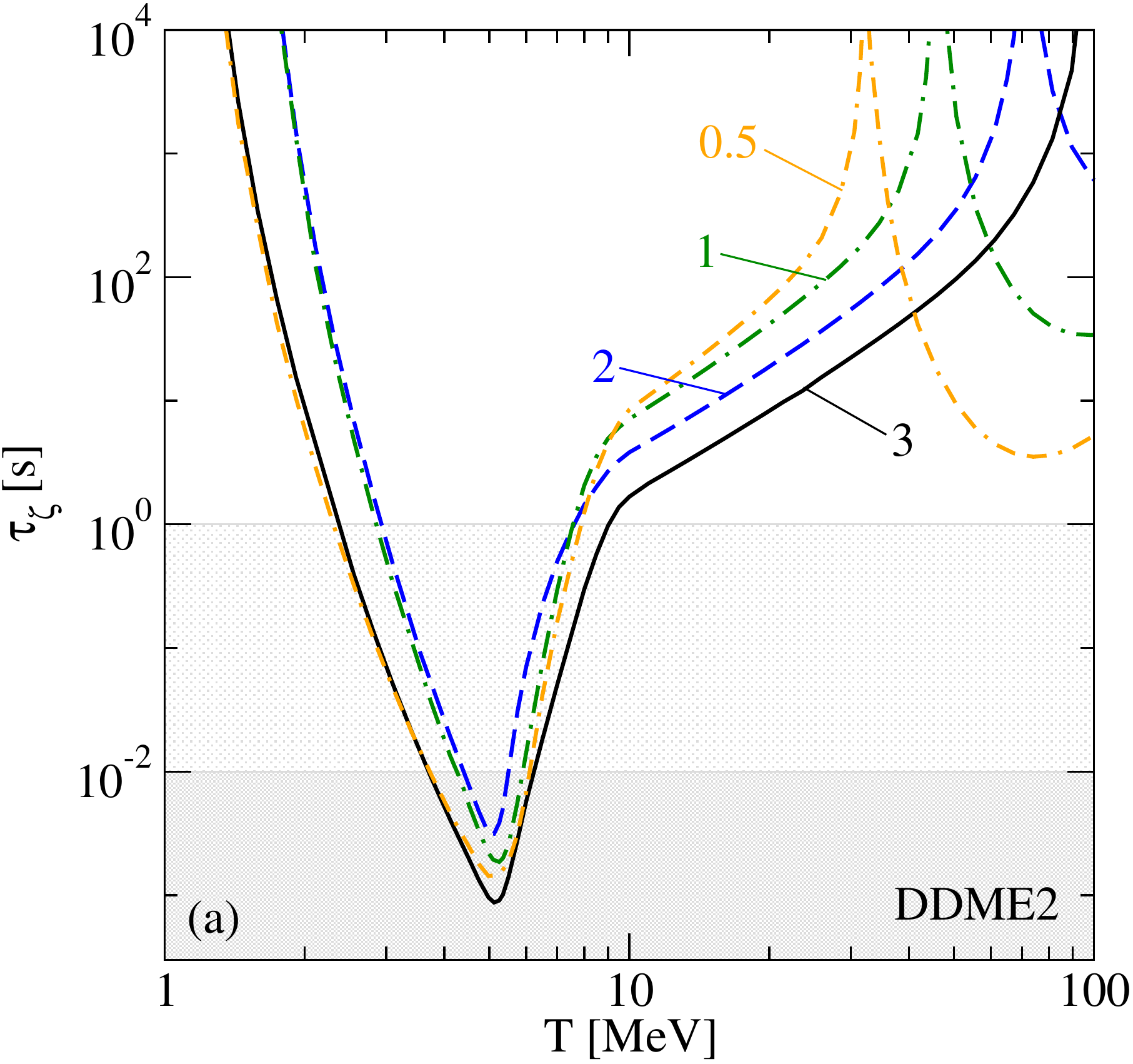}
\hskip 0.5cm
\includegraphics[width=0.45\columnwidth, keepaspectratio]{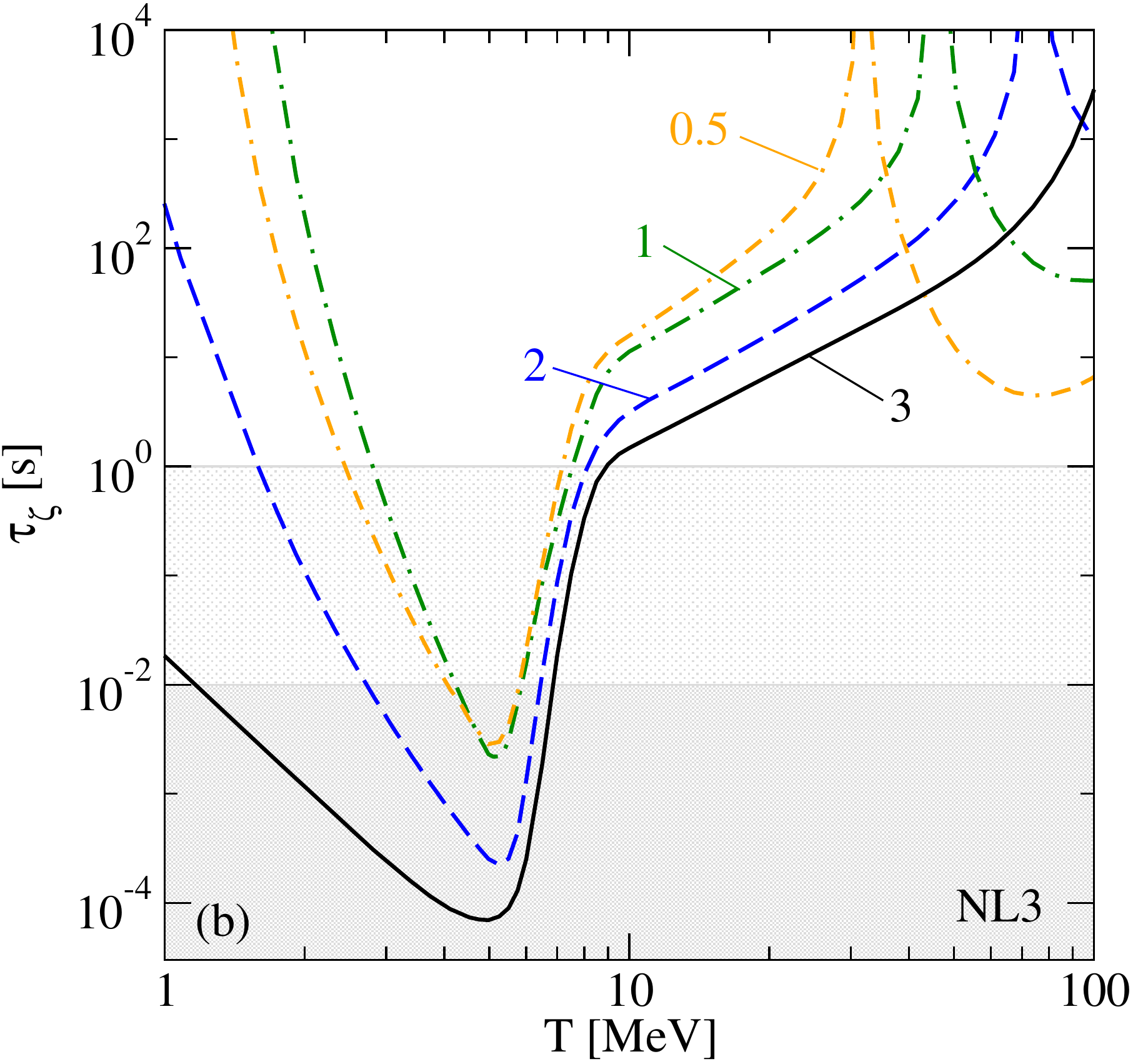}
\caption{The oscillation damping timescale as a function of temperature for various densities, interpolating between
the neutrino-transparent and neutrino-trapped 
regimes as in Fig.~\ref{fig:tau_damp1} but for $f=10$ kHz; 
(a) model DDME2; (b) model NL3.}
\label{fig:tau_damp10} 
\end{center}
\end{figure}

The gray shaded areas in Fig.~\ref{fig:tau_damp1} show the temperature regions  where the damping timescale is smaller than the characteristic timescales for the early  ($\simeq 10$\,ms, dark shaded areas) and long-term ($\simeq 1$ s, light shaded areas) post-merger evolution, respectively. It is seen that in massive neutron star mergers, bulk viscosity strongly damps density oscillations at densities $n_B\gtrsim n_0$ in the temperature range 3 $\lesssim T\lesssim $ 6 MeV for model DDME2 and 1 $\lesssim T\lesssim$ 6 MeV for model NL3. On the timescale of long-term evolution 
the damping is efficient also at lower densities, and the whole range of temperatures where the bulk viscosity  is relevant is 2 $\lesssim T\lesssim$ 7 MeV for 
model DDME2 and 1 $\leq T\leq$ 7 MeV for model NL3.

\begin{figure}[t] 
\begin{center}
\includegraphics[width=\columnwidth, keepaspectratio]{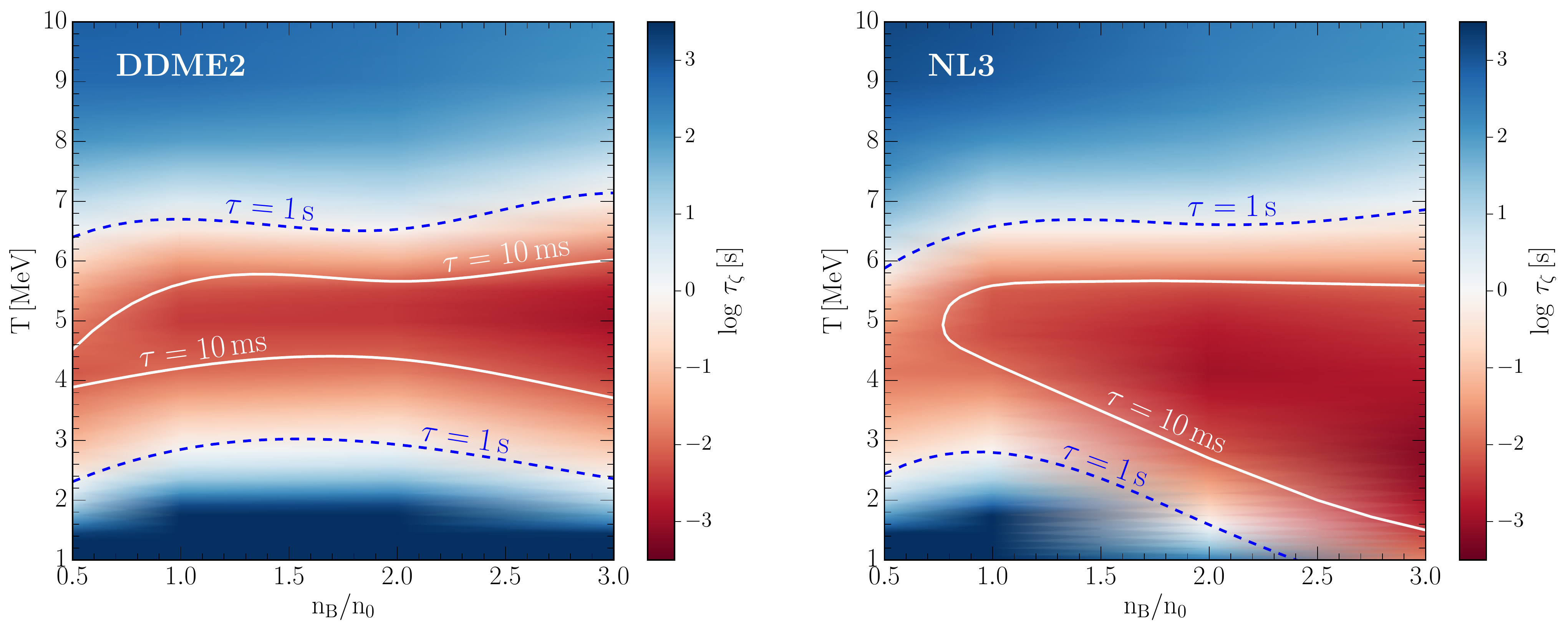}
\caption{ 
Bulk viscous damping timescale as a function of 
the density and temperature for model DDME2 (left panel) and model
NL3 (right panel). The density oscillation frequency is fixed at 
$f=1$ kHz. The white solid lines correspond to the characteristic timescale 
$\tau =10$\,ms, the blue dashed lines -
to the timescale $\tau =1$\,s. The areas colored in  red are the 
regions where the bulk viscous dissipation becomes important in 
the dynamics of neutron star mergers.
}
\label{fig:tau1_color} 
\end{center}
\end{figure}
\begin{figure}[!] 
\begin{center}
\includegraphics[width=\columnwidth, keepaspectratio]{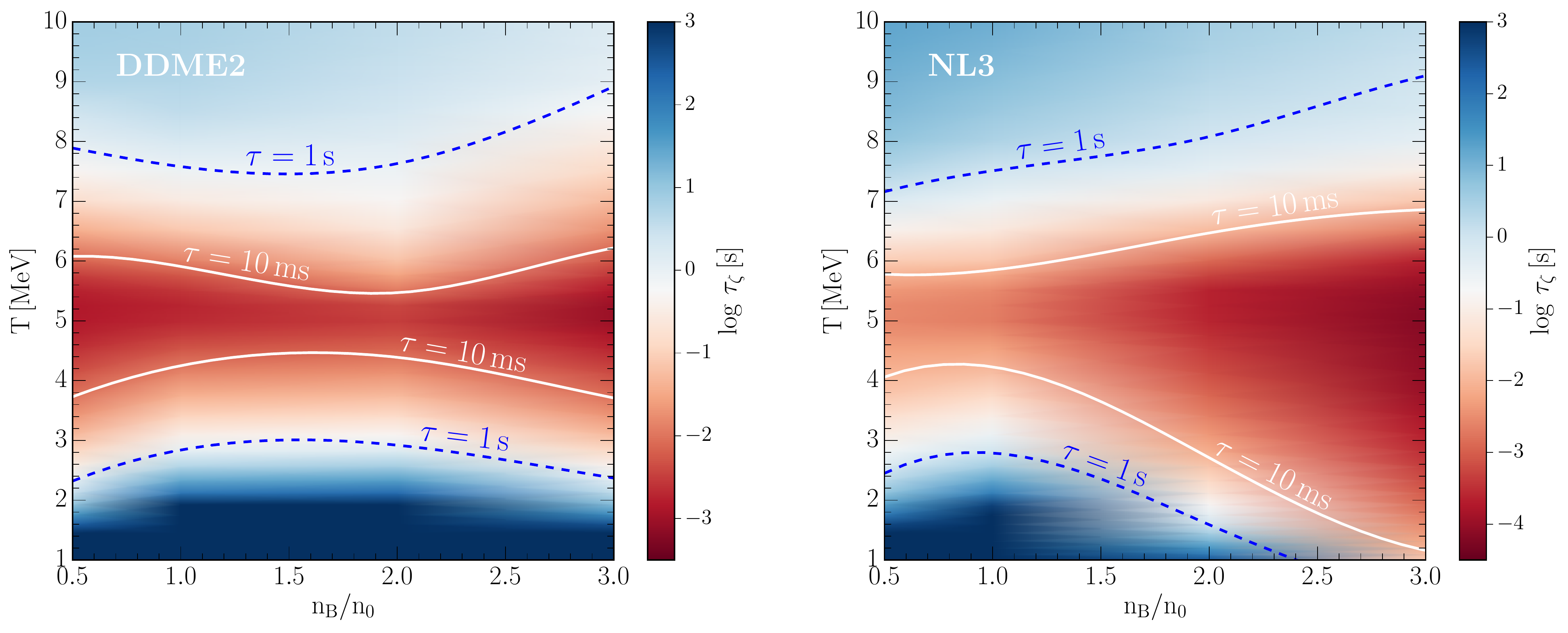}
\caption{ Same as Fig.~\ref{fig:tau1_color} but for frequency $f=10$ kHz.
}
\label{fig:tau10_color} 
\end{center}
\end{figure}

Comparing the two panels of Fig.~\ref{fig:tau_damp1}  we observe that at low temperatures $T\leq 4$ MeV and at densities $n_B\geq 2 n_0$ the damping timescales are much shorter for the EoS model NL3 which has a direct Urca threshold. Thus, the dynamics and observational signatures of post-merger objects potentially  contain information on whether the direct Urca processes are operative in the high-density domain of the neutron stars.

The temperature where the damping timescale reaches its minimum for 
$f=1$ kHz is located around $T\simeq 5$\,MeV at densities below the direct 
Urca threshold and around $T\simeq 3$\,MeV above the threshold.  
These results agree well with the results of Ref.~\cite{Alford2019a}
obtained within the Fermi-surface approximation. The exact computation,
however, obtained using exact beta-equilibrium condition for 
the neutrino-transparent matter
at finite temperatures~\cite{Alford2018b} suggests that the 
minimum of $\tau_\zeta$ is shifted to lower temperatures at 
densities which are below the direct Urca threshold, and,
as a result, the minimum always appears around $T\simeq 3$\,MeV~\cite{Alford2019a}
(note that the authors of Ref.~\cite{Alford2019a} included also
the modified Urca processes in their calculations, which, however,
do not change the location of the maximum of bulk viscosity as
their contribution is subdominant above $T=2 $\,MeV). 

The density-dependence of the damping timescale found here differs from that of Ref.~\cite{Alford2019a} where $\tau_\zeta$ was found to reach its minimum at low densities $n_B\lesssim n_0$. Apart from this, we find much lower values for the damping timescale at the minimum. This discrepancy arises because of the non-relativistic approximation for nucleon susceptibilities used in this work. This approximation works well at low densities, but strongly overestimates the susceptibility $C^2/A$ at higher densities $n_B\geq 2n_0$. 

The triangles in Fig.~\ref{fig:C2A_dens} show the values of the susceptibility $C^2/A$ obtained in 
Ref.~\cite{Alford2019a} for the models DD2 and IUFSU. We see that, 
although the relativistic corrections to the spectrum of nucleonic excitations
are about $20 \%$ at $n_B=2n_0$, they need to be included in the susceptibilities.
This will require also a fully relativistic study of the 
beta-equilibration rates which is relegated to a future work.

In Fig.~\ref{fig:tau_damp10}  we show the damping timescale for
10 kHz density oscillations. The minimum value of $\tau_\zeta$, in this case, is smaller than in the case of $f=1$ kHz by factors between 2 and 10, and the values of $\tau_\zeta$ in the neutrino-trapped regime are smaller by two orders of magnitude. However, the damping timescales of neutrino-trapped matter always remain larger than a second, since the bulk viscosity is not high enough to affect the evolution of mergers in this regime. 

Figures~\ref{fig:tau1_color} and 
\ref{fig:tau10_color} show the dependence of the bulk viscous damping timescale on the density and temperature colormap for the oscillation frequency fixed at $f=1$ kHz and $f=10$ kHz, respectively. The white solid and blue dashed lines show where the damping timescale becomes equal to the  timescales of 10 ms and 1 s,
respectively. In the areas shaded in dark red the bulk viscous damping timescale is $\tau_\zeta\leq 10$ ms, therefore, the  damping of density oscillations by the bulk viscosity is very efficient in those regimes. In the regions shaded in blue the 
role of the bulk viscosity in damping of oscillations is negligible, as the damping timescale $\tau_\zeta\geq 1$ s there.

For completeness we comment also on how our results will change if larger lepton fractions are considered. The case $Y_L=0.4$ was studied in our previous work Ref.~\cite{Alford2019b}, where the bulk viscosity was shown to be reduced 
by factors from 1 to 3 as compared to the $Y_L=0.1$ case. The pressure, and, therefore, also the nuclear incompressibility is only slightly sensitive to the lepton fraction. As a consequence, the damping timescales in 
the $Y_L=0.4$ case will be larger than in the $Y_L=0.1$ case by factors of a few, but the overall quantitative picture will remain the same.

In closing, we stress again that at densities $n_B\geq 2n_0$ the relativistic corrections to the spectrum of nucleonic excitations
become important for the computation of the bulk viscosity, and our results at high densities need to be improved accordingly. Also the appearance of hyperons and  other heavy baryons needs to be taken into account. Finally, we note that in the case of hybrid stars with quark cores, the bulk viscosity of quark matter can be important for damping of density oscillations (for computations in the case of cold compact stars see \cite{
Madsen1992,Drago2005,Alford:2006gy,Alford:2007rw,Manuel:2007pz,SaD2007b,SaD2007a,Alford:2008pb,Huang2010,Wang2010PhRvD}).

\section{Summary}
\label{sec:summary}
 
We have reviewed the computation, ingredients, and approximations involved in computations of bulk viscosity of nuclear matter at finite temperatures relevant to binary neutron star mergers. The bulk viscosity arises from the direct Urca $\beta$-equilibration reactions. A novel ingredient relative to the studies of cold neutron stars is the trapped neutrino component coexisting with the nuclear matter at temperatures $T\gtrsim 5$\,MeV.  The concrete computations were carried out with the relativistic density functional approach to the EoS of nuclear matter with two different parametrizations.

At a given value of oscillation frequency $\omega\equiv 2\pi f$ the bulk viscosity
shows the standard resonant form \eqref{eq:zeta}, with a maximum where the beta relaxation rate $\gamma$ matches $\omega$.  This resonant maximum is achieved in the temperature range where neutrinos escape from the merger region, since the relaxation rate at temperatures of a few MeV is sufficiently low to match the density oscillation frequency. The reason for lower relaxation rates as compared to the neutrino-trapped case is the suppression of the direct Urca processes at the relevant temperatures and densities.

When the temperature rises to the threshold for neutrino trapping  ($T\sim 5\,\MeV$) the bulk viscosity experiences a sharp fall by several orders of magnitude as the relaxation rate rises and the material enters the fast beta-equilibration regime with $\gamma{\gg} \omega$. In this regime the bulk viscosity is independent of the frequency and decreases with the temperature approximately as $\zeta\propto T^{-2}$. At temperatures of about $30$\,MeV a new feature appears: the bulk viscosity drops to zero at the temperature where the beta-disequilibrium--baryon-density susceptibility $C$ vanishes, and then rises again at higher temperatures.  The susceptibility vanishes because
the particle fractions become independent of the density and the material becomes scale-invariant.

The main new result of this work concerns the timescales of damping of density oscillations in neutron star mergers by the bulk viscous dissipation. As an input we used the results for the bulk viscosity in Ref.~\cite{Alford2019b}.  We find that the damping timescale has a minimum as a function of temperature, which is located at temperatures in the range 3$\div$6 MeV for various densities. Assuming oscillation frequency of 1\,kHz we find that the damping timescale at its minimum is of the order of ms, \ie, much shorter in the entire density range considered than the characteristic timescales of initial ($\sim 10$ms) and long-term ($\sim 1$s) post-merger evolution. We further find that the timescales of damping of density oscillations are shorter at the higher densities.  If the temperature is above the neutrino trapping temperature, the damping timescales are much longer 
as the bulk viscosity is strongly suppressed. Finally, we note that bulk viscous dissipation could be of interest in the context of hydrodynamics simulations of supernovas, where electron capture rates on protons and nuclei could be out of equilibrium (for recent numerical simulations, see \cite{Mezzacappa2015,Fischer2017,Connor2018ApJ,Burrows2020MNRAS}).

\vspace{6pt} 


\funding{The research of M. A. was funded by the U.S. Department of
  Energy, Office of Science, Office of Nuclear Physics under Award
  Number $\#$DE-FG02-05ER41375.  The research of A. H. and A. S. was
  funded by the Volkswagen Foundation (Hannover, Germany) grant
  No. 97029 and the European COST Action ``PHAROS'' (CA16214).  The
  research of A.S.\ was funded by Deutsche Forschungsgemeinschaft
  Grant No.  SE 1836/5-1. }

\acknowledgments{We thank Steven Harris, Kai Schwenzer, Alex Haber for
  discussions. M. A. and A. H. acknowledge the hospitality of Frankfurt Institute for Advanced Studies.}

\conflictsofinterest{The authors declare no conflict of interest.}

\reftitle{References}


\externalbibliography{yes}

\end{document}